\documentclass[10pt]{article}
\usepackage[utf8]{inputenc}
\usepackage[a4paper, total={6in, 8in}]{geometry}
\usepackage[T1]{fontenc}    
\usepackage{hyperref}       
\usepackage{url}            
\usepackage{booktabs}       
\usepackage{amsfonts}       
\usepackage{microtype}      
\usepackage{xcolor}         
\usepackage{amsmath,amsthm, amssymb}
\usepackage{graphicx}       
\usepackage{nicefrac}       
\usepackage{mathtools}
\usepackage{bm,verbatim} 
\usepackage{algorithm}
\usepackage{xspace}
\usepackage[noend]{algpseudocode}
\usepackage{multirow, multicol}
\usepackage{adjustbox}
\usepackage{pifont}
\usepackage{subcaption}
\usepackage{enumitem}
\usepackage[font=footnotesize,labelfont=bf]{caption}
\usepackage{tcolorbox}
\tcbuselibrary{skins,breakable}
\tcbset{enhanced jigsaw}
\newtheorem{mdalg}{Algorithm}

\theoremstyle{plain}
\newtheorem{theorem}{Theorem}[section]
\newtheorem{proposition}[theorem]{Proposition}
\newtheorem{lemma}[theorem]{Lemma}
\newtheorem{corollary}[theorem]{Corollary}
\theoremstyle{definition}

\theoremstyle{remark}

\usepackage[capitalize,noabbrev]{cleveref}
\crefname{property}{property}{Property}
\creflabelformat{property}{(#1)#2#3}
\crefname{equation}{eq}{Eq}
\creflabelformat{equation}{(#1)#2#3}

\DeclarePairedDelimiter{\norm}{\lVert}{\rVert}
\newcommand{\pqnorm}[1]{\norm{#1}_{p,q}}
\newcommand{\enorm}[1]{\norm{#1}_E}

\DeclareMathOperator{\Gram}{Gram}

\DeclareMathOperator{\Pow}{Pow}

\def\R{\mathbb{R}}

\makeatletter
\newcommand{\thickhline}{%
    \noalign {\ifnum 0=`}\fi \hrule height 1.2pt
    \futurelet \reserved@a \@xhline
} 
\newcommand*\samethanks[1][\value{footnote}]{\footnotemark[#1]}

\title{Johnson-Lindenstrauss Lemma Beyond Euclidean Geometry}
\author{Chengyuan Deng\thanks{Rutgers Uniersity, \texttt{\{cd751, jg1555, kll160, fluo, cx122\}@rutgers.edu}} \and Jie Gao\samethanks \and Kevin Lu\samethanks \and Feng Luo\samethanks \and Cheng Xin\samethanks}
\date{}

\newcommand{\disin}{{inner distance}}
\newcommand{\disout}{{cross distance}}
\newcommand{\normalizer}{{\mathrm{Norm}}}
\DeclareMathOperator{\diag}{\mathrm{Diag}}
\DeclareMathOperator{\dis}{\mathrm{DIS}}

\definecolor{azure}{rgb}{0.0, 0.5, 1.0}
\NewDocumentCommand{\cheng}{s m}{
    \IfBooleanTF{#1}
        {{\textcolor{azure}{#2}}} 
        {{\textcolor{azure}{Xin: #2}}} 
}

\newenvironment{tbox}{\begin{tcolorbox}[
		enlarge top by=5pt,
		enlarge bottom by=5pt,
		 breakable,
		 boxsep=0pt,
                  left=4pt,
                  right=4pt,
                  top=10pt,
                  arc=0pt,
                  boxrule=1pt,toprule=1pt,
                  colback=white
                  ]
	}
{\end{tcolorbox}}
\usepackage{amsmath}

\def\eqref#1{equation~\ref{#1}}

\def\1{\bm{1}}

\def\eps{{\varepsilon}}

\DeclareMathAlphabet{\mathsfit}{\encodingdefault}{\sfdefault}{m}{sl}
\SetMathAlphabet{\mathsfit}{bold}{\encodingdefault}{\sfdefault}{bx}{n}

\def\gN{{\mathcal{N}}}

\def\gX{{\mathcal{X}}}
\def\gY{{\mathcal{Y}}}

\def\sE{{\mathbb{E}}}

\def\sP{{\mathbb{P}}}

\makeatletter
\DeclareMathOperator*{\argmin}{\smash[b]{\operator@font arg\,min}}
\DeclareMathOperator*{\argmax}{\smash[b]{\operator@font arg\,max}}
\makeatother

\begin{document}
\maketitle

The Johnson-Lindenstrauss (JL) lemma is a cornerstone of dimensionality reduction in Euclidean space, but its applicability to non-Euclidean data has remained limited. This paper extends the JL lemma beyond Euclidean geometry to handle general dissimilarity matrices that are prevalent in real-world applications. We present two complementary approaches: First, we show the JL transform can be applied to vectors in pseudo-Euclidean space with signature $(p,q)$, providing theoretical guarantees that depend on the ratio of the $(p, q)$ norm and Euclidean norm of two vectors, measuring the deviation from Euclidean geometry. Second, we prove that any symmetric hollow dissimilarity matrix can be represented as a matrix of generalized power distances, with an additional parameter representing the uncertainty level within the data. In this representation, applying the JL transform yields multiplicative approximation with a controlled additive error term proportional to the deviation from Euclidean geometry. Our theoretical results provide fine-grained performance analysis based on the degree to which the input data deviates from Euclidean geometry, making practical and meaningful reduction in dimensionality accessible to a wider class of data. We validate our approaches on both synthetic and real-world datasets, demonstrating the effectiveness of extending the JL lemma to non-Euclidean settings.

\newpage

\section{Introduction}
\label{sec:intro}

The Johnson-Lindenstrauss (JL) lemma~\cite{johnson1984extensions} stands as a cornerstone result in dimensionality reduction. It states that random linear projection can reduce the dimensionality of datasets in Euclidean space, while approximately preserving pairwise distances. Formally, 

\begin{proposition}[Johnson-Lindenstrauss Lemma]
    \label{prop:jl}
    For any set of $n$ points $x_1, x_2, \dots x_n$ in $\mathbb{R}^d$ and $\eps \in (0,1)$, there exists a map $f:\mathbb{R}^d \rightarrow \mathbb{R}^m$, where $m = O(\log n/\eps^2)$ such that for any $i,j \in [n]$,
    \begin{equation}\label{eq:JL}
        (1-\eps)\|x_i-x_j\|_2 \leq \|f(x_i)-f(x_j)\|_2 \leq (1+\eps)\|x_i-x_j\|_2
    \end{equation}
\end{proposition}
In modern algorithm design, the JL lemma is widely used as a key component or pre-processing step for high-dimensional data analysis. In addition to effectiveness, another crucial reason for its significant impact is that the JL lemma can be achieved by random linear maps, which are data-independent and easy to implement. Johnson-Lindenstrauss Lemma has found numerous applications in machine learning, from the most immediate applications in approximating all pairs distances (when the input data is high-dimensional), to approximate nearest neighbor search~\cite{indyk07}, approximate linear regression~\cite{Pilanci2015-ah}, clustering~\cite{MakarychevMR19, izzo2021dimensionality,NarayananSIZ21,Becchetti19oblivious}, functional analysis~\cite{johnson2010johnson} and compressed sensing~\cite{baraniuk2008simple}. 

Note that two conditions must be met to apply the JL lemma: (1) the data points lie in high-dimensional \emph{Euclidean} space, and (2) the coordinates of all points are available. However, many real-world applications do not satisfy these two conditions. First, the dissimilarity measures used in modern data analysis are often non-Euclidean and sometimes even non-metric. Common examples include Minkowski distance, cosine similarity, Hamming distance, Jaccard index, Mahalanobis distance, Chebyshev distance, and Kullback–Leibler (KL) divergence, etc. Psychological studies have long observed that human similarity judgments do not conform to metric properties~\cite{Tversky1982-zt}. Second, high-dimensional coordinates may be unavailable or costly to obtain, whereas pairwise dissimilarities are easier to access. In recommendation systems, for instance, computing user or item embeddings can be expensive, while estimating pairwise dissimilarities (e.g. from co-click or co-purchase data) is relatively efficient. 
To date, our understanding of the JL lemma deviating from these two classical conditions mainly revolves around $\ell_p$ metrics and lower bound results.

In this paper, we study how to apply the Johnson-Lindenstrauss transform (a.k.a. random linear projection) in non-Euclidean, non-metric settings.
We provide theoretical analysis on the performance of the JL transform in these settings and show experiments with both synthetic and real-world data.

\paragraph{Our setting.} We consider the input as a dissimilarity matrix $D$ of size $n\times n$ where $D_{ij}$ is the dissimilarity between item $i$ and $j$. We make only two assumptions for the dissimilarity measure: symmetric ($D_{ij}=D_{ji}$) and reflexive ($D_{ii}=0$). Note that we do not require the triangle inequality to be satisfied, therefore the dissimilarity can be non-metric. In short, the input is a symmetric hollow dissimilarity matrix, and the expected output is a low-dimensional embedding of the original dataset that approximately preserves pairwise distances.

The major challenge we encounter in this setting is two-fold. The first is to obtain \emph{good} coordinates from a generic input dissimilarity matrix to represent the data, before we can even apply the JL transform. Second, we need a geometric characterization of the non-Euclidean non-metric setting, to show how different our setting stands from Euclidean geometry, which the JL Lemma is based on. 

Before we explain our results, we first mention existing lower bounds for dimension reduction in non-Euclidean settings. 
JL Lemma considers Euclidean spaces only. Naturally, one asks whether such a result is possible for other spaces. There are several non-trivial dimension reduction results for $\ell_1$ norm~\cite{Matousek2008-qa} and $\ell_p$ norm~\cite{Lee2005-nc}. However, the target embedding dimension for $\ell_1$, $\ell_p$ and nuclear norm is necessarily polynomial in $n$ for constant distortion~\cite{Schechtman2011-tw, Naor2020-vg, Brinkman2005-uq}. These lower bounds remind us that a worst-case guarantee on low distortion and logarithmic target dimension is not possible. Rather, our results provide error analysis which depends on parameters characterizing how the input data deviates from Euclidean geometry. In other words, we have fine-grained performance analysis based on the introduced parameters.

\paragraph{Our Contributions.} We present two approaches to recover coordinates that fit into the non-Euclidean non-metric setting. For both approaches, we generalize the Euclidean norm $\ell_2$ to a new form to capture the input data geometry, together with certain geometric parameters indicating how far it deviates from Euclidean geometry. We give error analysis on the dissimilarity distortion, which may involve an additive term whose magnitude is proportional to the geometric parameters.

At the heart of our first approach is the observation that any symmetric hollow matrix can be written as the distance matrix of vectors in \emph{pseudo-Euclidean space}~\cite{GOLDFARB1984575,Pekalska10.5555/1197035,deng24NeucMDS}. Here the distance between two vectors $x=(x_1, x_2,... x_n)$ and $y=(y_1, y_2,... y_n)$ is captured by a bilinear form of signature $(p,q)$, which is defined as $\langle x, y\rangle_{p,q} = \sum_{i=1}^p x_iy_i - \sum_{i=p+1}^{p+q}x_iy_i$. The squared $(p, q)$-distance between $x$ and $y$ is $\|x-y\|^2_{p, q}=\langle x-y, x-y\rangle$. When $p=n, q=0$, it is the squared Euclidean norm. We will give a more detailed introduction later. At this point, it suffices to interpret the parameter $p$ resembling and $q$ negating the Euclidean space. Our first result shows the generalization of the JL lemma to the pseudo-Euclidean geometry. 

\begin{theorem}[Fine-grained JL lemma, informal Version of \Cref{thm:jl-pq}]
   Given any symmetric hollow dissimilarity matrix $D$ of size $n$, we are able to obtain embeddings $X$ in the pseudo-Euclidean space. For any $\varepsilon \in (0,1)$, there exists a JL transform to $X$ with target dimension $O(\log n/\varepsilon^2)$, such that any pairwise dissimilarity $D_{ij}$ is preserved with at most $1 \pm \varepsilon \cdot C_{ij}$ multiplicative factor, where $C_{ij}$ is the ratio of the squared Euclidean distance and squared $(p, q)$ distance. 
\end{theorem}

Our second result takes a different route. We prove that a symmetric hollow matrix is also a matrix of \emph{generalized power distances}. Given two points $x$ and $y$ with respective radius $r_x$ and $r_y$, the generalized power distance is defined as $\|x-y\|^2_E -(r_x+r_y)^2$, where $\|x-y\|^2_E$ denotes the Euclidean norm. This is the squared length of the tangent line segments of two balls centered at $x, y$ with radii $r_x, r_y$. Again, at this point, it suffices to interpret $r_x, r_y$ as a measure of deviation from Euclidean space, with both a geometric meaning and a statistical interpretation of distance between points with uncertainties. 

We show that any input symmetric hollow matrix $D$ of size $n$ can be written as the generalized power distance matrix of $n$ points $\{p_i\}$ with the same radius $r=\sqrt{|e_n|}/2$, where $e_n$ is the smallest eigenvalue of the Gram matrix of $D$. This linear algebra result may be of independent interest. Only when $D$ is a Euclidean distance matrix, all eigenvalues are non-negative and $r=0$. 
Our second result follows from applying the JL transform on the ball centers (i.e. $\{p_i\}$). We obtain a $(1 \pm \varepsilon)$ multiplicative approximation of the generalized power distance with an additive error of $4\varepsilon r^2 $. 


\begin{theorem}[Power-distance JL Lemma, informal Version of \Cref{thm:JL-power}]
\label{thm:infml-jl-power}
   Given any symmetric hollow dissimilarity matrix $D$ of size $n$ and $\varepsilon \in (0,1)$, there exists a JL transform with target dimension $m=O(\log n/\varepsilon^2)$, such that any pairwise distance $D_{ij}$ is preserved with at most $1\pm \varepsilon$ multiplicative factor and an $4\varepsilon r^2$ additive factor, where $r=\sqrt{|e_n|}/2$ with $e_n$ as the smallest eigenvalue of the Gram matrix of $D$.
\end{theorem}

To complement \Cref{thm:infml-jl-power}, we are able to extend the techniques in~\cite{Larsen17optimality} and give a lower bound of $\Omega(\log n/\varepsilon^2)$ on the target dimension to achieve the multiplicative and additive factors as mentioned. Note that $m = \Omega(\log n/\varepsilon^2)$ matches the JL lemma lower bound.

\textbf{Experiments.} We implement both methods of JL transform with respect to the above results and evaluate their performances on $10$ datasets. We observe that the experiment results corroborate with our theoretical results, and outperform classical JL transform consistently on non-Euclidean datasts. 
Our codes are on the Anonymous Github\footnote{\url{https://anonymous.4open.science/r/Non-Euclidean-Johnson-Lindenstrauss-1673}}.

\subsection{Related Work}

\textbf{Random Linear Projection and JL Lemma} 
A nice survey on JL transform can be found in~\cite{Nelson2020-dm}. 
There was a series of work~\cite{Alon2003-iw,Kane2011-un,Jayram2013-ti,larsen_et_al:LIPIcs.ICALP.2016.82} that studied whether the bound on the target embedding dimension in JL lemma is tight, starting from the original JL paper~\cite{johnson1984extensions}, and the optimality of JL lemma for Euclidean dimension reduction is finally established even among non-linear embeddings~\cite{Larsen17optimality,Alon2017-yf}.
In terms of algorithms, there have been developments of variants of random projections that are more friendly for implementations 
\cite{Achlioptas2003-pg,Ailon2009-of,Dasgupta10.1145/1806689.1806737,yin2020extremely,Kane2014-rn}.
Empirical study of JL Lemma can be found in~\cite{Venkatasubramanian2011-jf}.


Last, Benjamini and Makarychev~\cite{Benjamini2008-hh} considered dimension reduction using the Poincar\'e half-space 
model of hyperbolic space $\mathbb{H}^n$ where each point is represented by $(x, z)$ with $x \in \mathbb{R}^{n-1}$ and  height $z \in \mathbb{R}^+$.
They considered using JL random projection $f$ on the Euclidean part and obtain $(f (x), z)$ with $f(x)$
in Euclidean space of dimension $O(\log n/\eps^2)$, achieving $1 + \eps$ distortion.

\textbf{Data-dependent dimension reduction.} A number of dimension reduction techniques that are also popular in practice include Multidimensional scaling (MDS)~\cite{Torgerson1952-bz} and Principal Component Analysis (PCA)~\cite{Pearson1901-xe}. These methods are \emph{data-dependent} and choose linear projections based on the input data. There has also been recent work trying to apply these methods on non-Euclidean data, for example PCA in hyperbolic or spherical geometry~\cite{GuSGR19,chami21horoPCA,Tabaghi2024-bn,Fan2022-mx}.
Manifold learning (such as Isomap~\cite{Tenenbaum2000-ta} and LLE~\cite{Roweis2000-an}) considers input data points that come from a low-dimensional manifold, recovers the distances between points along the manifold and applies existing dimension reduction methods.  
A recent work called Non-Euclidean MDS~\cite{deng24NeucMDS} considers dimension reduction of pseudo-Euclidean settings and minimizes the loss called STRESS (the sum of
squared difference of pairwise embedding distances to the input dissimilarities). It does not provide error analysis or guarantees for individual dissimilarity distortion. 




\section{Johnson-Lindenstrauss Lemma in Pseudo-Euclidean space}\label{sec:jl-pq}

\subsection{Pseudo-Euclidean space of signature $(p, q)$} \label{subsec:pq}

We consider the $d$-dimensional vector space $\R^d$ equipped with a $(p, q)$ bilinear form $\langle,\rangle$ where $p+q=d$.  For simplicity, we use $\R^{p,q}$ to denote $\R^d$ with the $(p,q)$ bilinear form $\langle,\rangle$. 
Consider two $d$-dimensional (column) vectors $u, v\in \R^{p,q}$, with $u=(u_1, u_2,... u_{p+q})$ and $v=(v_1, v_2,... v_{p+q})$, we define the bilinear norm as: 
\[\langle u, v\rangle = \sum_{i=1}^p u_iv_i - \sum_{i=p+1}^{p+q}u_iv_i\]
Equivalently, we can write $\langle u, v\rangle = u^T \Lambda v$ with $\Lambda$ as a $d\times d$ diagonal matrix\footnote{In general, for the bilinear form $\Lambda$, there can also be additional $r$ dimensions with zero as diagonal elements, in addition to $p$ elements of $1$ and $q$ elements of $-1$. In that case $p+q+r=d$. In our writing we skipped $r$ as they do not contribute to the similarity values for ease of exposition. } with the first $p$ diagonal elements as $1$ and the remaining diagonal elements as $-1$. 
For a vector $u$, its \emph{scalar square} is $\langle u, u\rangle$. For two vectors $u, v$ their \emph{interval square} is obtained by $\langle u-v, u-v\rangle$.

Consider a given symmetric dissimilarity matrix $D \in \mathbb{R}^{n \times n}$, where $D_{ij}$ refers to the dissimilarity measure of element $i$ and element $j$. We assume that $D$ is hallow, $D_{ii}=0$, and symmetric, $D_{ij}=D_{ji}$. The following proposition shows that a set of vectors in the Pseudo-Euclidean space of signature $(p,q)$ can be obtained via $D$.
\begin{proposition}
     For any hollow symmetric matrix $D$, there is a symmetric bilinear form $\langle,\rangle_{p,q}$ for integers $p,q$ and $n$ vectors $x_1, \dots, x_n$ such that $D_{ij} = \langle x_i-x_j, x_i-x_j\rangle$.
\end{proposition}
We now explain how to obtain $p,q$ and vectors $x_i$. With an input $D$, we first perform centralization to obtain the Gram matrix $B=\Gram(D)=-CDC/2$, 
where $C = I - \frac{1}{n} \mathbf{1}_n\mathbf{1}_n^T$ is the centering matrix and $\mathbf{1}_n$ is a vector of ones.
Since $-CDC/2$ is a symmetric matrix, its eigenvalues are real, denoted as $\lambda_1 \geq \lambda_2 \geq \cdots \geq \lambda_n$. We take $p$ as the number of non-negative eigenvalues and $q$ to be the number of negative eigenvalues. $p+q=n$. Further, suppose $U \in \mathbb{R}^{n \times n}$ is the orthogonal vectors, we have
\begin{equation*}\label[eq]{eq:eig_decomp}
   B= -CDC/2 = U \diag(\lambda_1, \cdots, \lambda_n) U^T
\end{equation*}
Now we can recover the coordinates of the $n$ elements as $n$-dimensional vectors. Specifically $X$ is a matrix of dimension $n\times n$ with the columns representing the coordinate vectors of the $n$ points.
\begin{equation*}\label[eq]{eq:coordinates}
X=(x_1, \cdots, x_n)= \diag(\sqrt{|\lambda_1|}, \cdots, \sqrt{|\lambda_n|})  \cdot U^T
\end{equation*}
This way, we have $B= -CDC/2 = X^T\Lambda X$, with $\Lambda$ as an $n\times n$ diagonal matrix with the first $p$ diagonal elements as either $1$ (or $0$) if the corresponding eigenvalue is positive (or $0$), and the remaining diagonal elements as $-1$. Equivalently, we have \begin{equation*}\label[eq]{eq:distance}
D_{ij}=(x_i-x_j)^T\Lambda (x_i-x_j)=\langle x_i-x_j, x_i-x_j \rangle.
\end{equation*}
That is, 
the dissimilarity $D_{ij}$ is precisely the $(p, q)$ interval square of $x_i, x_j$.

If the input dissimilarity matrix $D$ is the squared Euclidean distance matrix of $n$ points, the Gram matrix is positive semi-definite and the above procedure would recover the Euclidean coordinates $\{x_i\}$. This is referred to as the classical Multidimensional scaling (MDS)~\cite{Torgerson1952-bz}. If the input dissimilarity matrix $D$ generalizes beyond the squared Euclidean distance matrix but remains a hallow symmetric matrix, we can still recover the coordinates so that they produce the entries in $D$ using a general $(p, q)$ bilinear form. Notice that the above procedure generates coordinate vectors in dimension $n$. 

The $(p, q)$-space has deep connections to spacetime and special relativity theory. Specifically, the Lorentzian $n$-space is the vector space $\mathbb{R}^n$ with the $n$-dimensional Lorentzian inner product, where the squared norm of a vector $u=(u_0, u_1, \cdots, u_{n-1})$ has the form $|u|=-u^2_0 + u_1^2+\cdots + u^2_{n-1}$. 
Thus the Loerentizan space has signature $(n-1, 1)$.
Four-dimensional Lorentzian space is called the \emph{Minkowski space} and forms the basis of special relativity. Furthermore, the collection of vectors in $\mathbb{R}^{n+1}$ with Lorentzian inner product of $-1$ has imaginary Lorentzian length and is precisely the hyperboloid model of the hyperbolic $n$-space $\mathbb{H}^n$~\cite{Ratcliffe2006-ml}. 

\subsection{Johnson-Lindenstrauss Lemma in $(p, q)$-space}

We start with some notation. For a vector $v=(v_1, v_2, \cdots, v_{p+q})\in \mathbb{R}^{p,q}$, we will use $\pqnorm{v}^2$ to denote the "$(p, q)$-norm" $\langle v, v\rangle$ and $\enorm{v}^2$ to denote the squared Euclidean norm of $v$. Further, we denote $v^{(p)}=(v_1, v_2\cdots v_p)\in \mathbb{R}^p$ and $v^{(q)}=(v_{p+1}, v_{p+2}, \cdots, v_{p+q})\in \mathbb{R}^q$. 
Then $\pqnorm{v}^2=\enorm{v^{(p)}}^2-\enorm{v^{(q)}}^2$
and $\enorm{v}^2=\enorm{v^{(p)}}^2+\enorm{v^{(q)}}^2$.

First, we will prove a Johnson-Lindenstrauss style dimension reduction by applying the JL-lemma  on the $p$ part and $q$ part respectively. The proofs in this section can be found in \Cref{sec:proofs}.

\begin{lemma}\label{lem:jl-pq}
       For any set of $n$ points $x_1, x_2, \dots x_n$ in $\mathbb{R}^{p, q}$ and $\eps \in (0,1)$, there exists a map $f:\mathbb{R}^{p, q} \rightarrow \mathbb{R}^{p', q'}$, where $p', q' = O(\log n/\eps^2)$ such that for any $i,j \in [n]$,
    \begin{equation}\label{eq:JL-pq}
     \pqnorm{x_i-x_j}^2-\eps \enorm{x_i-x_j}^2 \leq \|f(x_i)-f(x_j)\|_{p' ,q'}^2 \leq \pqnorm{x_i-x_j}^2+\eps \enorm{x_i-x_j}^2
    \end{equation}
\end{lemma}

Essentially for $(p, q)$-space we can still have a similar JL-lemma, which, compared with the standard JL-lemma~\Cref{eq:JL}, we do not have the $1+\eps$ multiplicative error but rather an additive error to the squared Euclidean norm. This gives us an immediate JL-style result when $\pqnorm{x_i-x_j}^2$ and $\enorm{x_i-x_j}^2$ are a constant factor of each other. Formally, we have the following statement.
\begin{theorem}\label{thm:jl-pq}
    For any set of $n$ points $x_1, x_2, \dots x_n$ in $\mathbb{R}^{p, q}$ and $\eps \in (0,1)$, there exists a map $f:\mathbb{R}^{p, q} \rightarrow \mathbb{R}^{p', q'}$, where $p', q' = O(\log n/\eps^2)$ such that for any $i,j \in [n]$,
    \begin{equation}\label{eq:JL-pq}
     (1-\eps \cdot C_{ij})\pqnorm{x_i-x_j}^2 \leq \|f(x_i)-f(x_j)\|^2_{p',q'} \leq (1+\eps \cdot C_{ij})\pqnorm{x_i-x_j}^2,
    \end{equation}
    where $C_{ij}=\left |\frac{\enorm{x_i-x_j}^2}{\pqnorm{x_i-x_j}^2}\right |$.
\end{theorem}
A few remarks are in place. First, one can apply any JL-style algorithm for the $p$ and $q$ parts, for example, by using the fast JL algorithm. Suppose $f_p$ is implemented by a random matrix $M_{p'p}$ and $f_q$ is implemented by a random matrix $M_{q'q}$. Then $f$ is implemented by a matrix $M$ of dimension $(p'+q')\times (p+q)$ with $M_p$ and $M_q$ at the diagonal and zero elsewhere, $f(x)=Mx$.
Second, random projection for the $(p, q)$ space setting provides an error-bound guarantee for a pair of points that gracefully degrades as the $(p, q)$ norm deviates from the Euclidean norm. 
Below we show a few natural situations where the $(p, q)$ norm and Euclidean norm are indeed bounded. 


\begin{lemma}\label{lem:highp-lowq}
Let $v \in \mathbb{R}^{p,q}$. If each coordinate of $v$ is selected from the same distribution, $f$ with a finite second moment, then we have with high probability, $\enorm{v}^2 < C\pqnorm{v}^2$ if $q < \frac{C-1}{C+1} p$ or $\enorm{v}^2 < -C\pqnorm{v}^2$ if $p < \frac{C-1}{C+1} q$ for $C > 1$. 
\end{lemma}

\begin{corollary}
A random vector $v$ chosen uniformly over the unit sphere $\mathbb S^{p+q-1}$ has with high probability, $\enorm{v}^2 < C\pqnorm{v}^2$ if $q < \frac{C-1}{C+1} p$.
\end{corollary}

Now we put together all of the previous results into a single statement on a set of random points in $(p, q)$ space.

\begin{theorem}\label{thm:random-pts-pq}
Let $X \subset \mathbb{R}^{p, q}$ be a randomly selected point set of size $n$, with $q < \frac{C-1}{C+1} p$ for a constant $C$. For all $u, v \in X$, $\frac{u-v}{\enorm{u-v}}$ has distribution equal to uniformly choosing a vector on $S^{p+q-1}$.
Then, with high probability, there exists a map $f:\mathbb{R}^{p, q} \rightarrow \mathbb{R}^{p', q'}$, $p', q' = O(\frac{C^2\log(n)}{\epsilon^2})$ such that for any $i,j \in [n]$,
\begin{equation}\label{eq:JL-pq}
 (1-\epsilon)\pqnorm{x_i-x_j}^2 \leq \|f(x_i)-f(x_j)\|_{p' ,q'}^2 \leq (1+\epsilon)\pqnorm{x_i-x_j}^2 .
\end{equation}
\end{theorem}




\section{Johnson-Lindenstrauss Lemma for Generalized Power Distances}
\label{sec:jl-power}

\subsection{Generalized Power Distance}

\begin{figure}[htbp]
\centering
\includegraphics[width=0.3\linewidth]{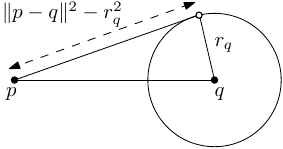} \hspace*{5mm}
\includegraphics[width=0.35\linewidth]{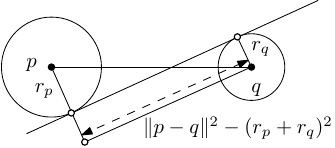} 
\includegraphics[width=0.18\linewidth]{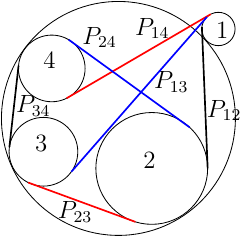} 
\caption{Left: power distance from a point $p$ to a ball at $q$ of radius $r_q$; Middle: power distance between two balls at points $p, q$ with radius $r_p$ and $r_q$ respectively; Right: Casey's Theorem.}
\vspace{-4mm}
\label{fig:power}
\end{figure}

Given two balls centered at $p, q\in \mathbb{R}^d$, with radii $r_p, r_q$, we define the \emph{generalized power distance} as:
\begin{equation}\label{poerdis}
    \Pow((p, r_p), (q, r_q) ) = \|p-q\|^2_E - (r_p + r_q)^2.
\end{equation}
This power distance measures the distance of the internal tangents between two disjoint circles (the tangent line that keeps two circles on different sides. See the middle picture in Figure \ref{fig:power}).
Note that there is another second generalized power distance between two circles centered at $p_i$ of radius $r_i$ given by $||p_1-p_2||^2_E -(r_1-r_2)^2$, which measures the distance of the external tangents between the circles (the tangent line that keeps two circles on the same side).

\textbf{Geometric Interpretation.}
In elementary plane geometry, the power of a point is a real number that reflects the relative distance of a given point from a given circle, introduced by Jakob Steiner in 1826. The power of a point $p$ with respect to a circle with center $q$ and radius $r$ is defined as $\|p-q\|^2-r^2$. This value is positive when $p$ is outside the circle, and by Pythagorean theorem it is squared tangential distance $|pt|$ where $t$ is a tangent point of $p$ to the circle. When $p$ is inside the circle, the power is negative, and precisely the negative of squared distance $|ps|$ where $s$ is a point on the circle with $\triangle psq$ to be a right triangle with $\angle spq$ as 90 degree. 

The generalized power distance considers the distance between two circles and reduces to the classical power distance if one of the circles is a point. 
Furthermore, two circles are disjoint if and only if their generalized power distance is positive.  In this case, their generalized power distance is equal to the square of the distance
between the two tangent points on the common internal tangent line between the two circles (with the two circles on different sides).  
See \Cref{fig:power} for examples. The generalized power distance can be called power distance of \emph{weighted points} with $r_p$ as the weight of $p$.

There are several interesting properties of the generalized power distance. 
First, the Casey's theorem~\cite{Casey1864,Emch1917-rk} still holds for generalized power distance. 
That is, if four circles in the plane are tangent to the fifth circle, then the generalized power distances $P_{ij}$, $i, j\in \{1, 2, 3, 4\}$, between them satisfy the Ptolemy identity $P_{12}P_{34}+P_{23}P_{41}=P_{13}P_{24}$.  This Ptolemy identity plays a key role in applications of the generalized power distances. Indeed, the Ptolemy identity and the associated Ptolemy inequality $P_{12}P_{34}+P_{23}P_{41}\geq P_{13}P_{24}$ have been used in optimization and approximation for prune search space and optimize geometric networks, for approximate distance estimate, circle fitting and cyclic polygon detection; in distance geometry for ensuring distance constraints for sensor networks~\cite{Khadivi2010-db} and molecule reconstruction~\cite{distance}.
In metric geometry~\cite{Burago01}, a metric space is called Ptolemaic space if Ptolemy inequality holds for any quadriple points.  It is well known that Euclidean space and any inner-product spaces are Ptolemaic and a normed vector space is Ptolemaic if and only if the norm comes from an inner product~\cite{Schoenberg1940-ss,Schoenberg1952-po}. Furthermore, a Riemannian manifold is a Ptolemaic space if and only if its sectional curvature is non-positive ~\cite{hadama}. 
It shows that Ptolemy inequalities can be used for studying curvature properties of metric spaces.  Recall that a length space $(X, d)$ is Gromov $\delta$-hyperbolic if it satisfies  Gromov’s Four-Point Condition:
for any $x,y,z,w \in X$, $d(x,z)+d(y,w) \leq \max\{d(x,y)+d(z,w), d(x,w)+d(y,z)\}+2\delta$. The Ptolemy inequality is another four-point condition on metric spaces for investigating curvature of general spaces~\cite{Foertsch2010-uw}.



\textbf{Any Symmetric Dissimilarities are Power Distances.}
Now, we show that any symmetric hollow dissimilarity matrix can be written as the matrix of power distances between $n$ weighted points.
\begin{lemma}\label{lem:power-D}
    Given any $n\times n$ symmetric hollow dissimilarity matrix, $D$, we can rewrite $D = E + 4r^2(I - J)$ where $r \in \mathbb{R}^+$, $E$ is a Euclidean distance matrix, $I$ is an $n\times n$ identity matrix, $J = \mathbf{1}_n\mathbf{1}_n^T$. More specifically $E$ is a Euclidean distance matrix if and only if $2r^2 \geq |e_n|$ where $e_n$ is the least eigenvalue of $\Gram(D)$.
\end{lemma}

From \Cref{lem:power-D}, suppose $E$ is an Euclidean distance matrix where the $(i, j)$ element is $\|p_i-p_j\|^2$, with $n$ points $\{p_i\}$. Then $D$ is the matrix of generalized power distances $\{(p_i, r)\}$, i.e., with the $(i,j)$ element as $\|p_i-p_j\|^2_E-4r^2$.
Here all the weighted points have radius $r$ which can be seen as a measure of how far the input matrix $D$ is away from being Euclidean. 
When $D$ is a Euclidean distance matrix, the Gram matrix $\Gram(D)$ is positive semi-definite and the smallest eigenvalue is $e_n=0$. Thus $r=0$ and the generalized power distance reduces to the squared Euclidean distance. When $D$ is no longer a Euclidean distance matrix, $\Gram(D)$ necessarily have negative eigenvalues, i.e., $e_n<0$. This makes $r$ to be non-trivial.   

\textbf{Statistical Interpretation} The generalized power distance also has a statistical interpretation using the silhouette coefficient~\cite{Silhouettes_Rousseeuw1987}, which measures how similar two clusters are. Consider two clusters $X, Y$, the \emph{silhouette value} of a point $x\in X$ is the average distance from $x$ to points in $Y$ minus the average distance from $x$ to other points in $X$. The \emph{silhouette coefficient} of the two clusters $X, Y$ is the average of the silhouette coefficient of all points in $X, Y$. 
In \Cref{appendix:silhouette} we show that the silhouette coefficient of 
two Gaussian distributions $\gX=\gN(\mu_x, \sigma_x), \gY=\gN(\mu_y, \sigma_y)$ is 
$SC_W(\gX, \gY) = \|\mu_x- \mu_y\|^2- (\sigma_x + \sigma_y)^2.$ Notice that this is precisely the generalized power distance of weighted points $(\mu_x, \sigma_x)$ and $(\mu_y, \sigma_y)$. 

Therefore, \Cref{lem:power-D} suggests that any input dissimilarity matrix $D$ that is hollow and symmetric is actually the silhouette coefficients of $n$ Gaussian distributions centered at $\{p_i\}$ with variance $r$ which is given by \Cref{lem:power-D}. 
Note that this statistical interpretation resonates with prior literature~\cite{Xu2011-fo,Duin2010-wh} which identified that one major source of data in non-Euclidean geometry is due to measurement noises and uncertainties. 



\subsection{Johnson-Lindenstrauss Lemma for Power Distance}

\begin{lemma}\label{lem:JL-power}
    Given $n$ weighted points $(p_i, r_i)$ with $p_i\in \mathbb{R}^d$, there exists a map $f:\mathbb{R}^d \rightarrow \mathbb{R}^m$, where $m = O(\log n/\eps^2)$ such that for any $i\neq j \in [n]$,
    \begin{align}\label{eq:power-jl}
        (1-\eps)\Pow((p_i, r_i), (p_j, r_j))-\eps (r_i+r_j)^2 &\leq \Pow((f(p_i), r_i), (f(p_j), r_j)) \\
        &\leq (1+\eps)\Pow((p_i, r_i), (p_j, r_j))+\eps (r_i+r_j)^2.
        \end{align}
\end{lemma}

With Lemma \ref{lem:power-D} and Lemma \ref{lem:JL-power} we immediately have the following.

\begin{theorem}
\label{thm:JL-power}
Given any $n\times n$ symmetric hollow dissimilarity matrix, $D$, with power distance representation by $\{(p_i, r)\}$ where $r= \sqrt{|e_n|}/2$ and $e_n$ is the smallest eigenvalue of $\Gram(D)=-\frac{1}{2}CDC$ with $C=1-J/n$ and $J$ is an all $1$ matrix, there exists a map $f:\mathbb{R}^n \rightarrow \mathbb{R}^m$, where $m = O(\log n/\eps^2)$ such that for any $i\neq j \in [n]$,
    \begin{align*}
        (1-\eps)\Pow((p_i, r), (p_j, r))-\eps 4r^2 &\leq \Pow((f(p_i), r), (f(p_j), r)) \\
        &\leq (1+\eps)\Pow((p_i, r), (p_j, r))+\eps 4r^2.
    \end{align*}
\end{theorem}

Last, we remark that the error bounds in Lemma \ref{lem:JL-power} also imply a $1+\eps$ distortion for dimension reduction on the Euclidean distances of the centers (as in the JL Lemma~\ref{eq:JL}). Therefore, the lower bound on the target dimension being $\Omega(\log n/\eps^2)$ for a dimension reduction algorithm satisfying the error bounds in Lemma \ref{lem:JL-power} for generalized power distances holds, by the same argument in the proof of dimension optimality of standard JL-lemma~\cite{Larsen17optimality}.

\section{Non-Euclidean Johnson-Lindenstrauss Transforms}
\label{sec:algo}


We present non-Euclidean JL transforms, i.e. the algorithmic results corresponding to \Cref{thm:jl-pq} and \Cref{thm:JL-power}.  In both cases we employ classical JL transform as a standard technique, and we assume an input dissimilarity matrix $D$ which is symmetric ($D_{ij}=D_{ji}$) and hollow ($D_{ii}=0$), in addition to the standard distortion factor $\eps \in (0,1)$.
\begin{tbox}
    \textbf{Pseudo-Euclidean JL Transform}
    \begin{enumerate}
    \item 
    Find the orthogonal decomposition of the dissimilarity matrix $D = O\Lambda O^T$ where $\Lambda$ is a diagonal matrix of the eigenvalues.
    \item 
    Let $A$ be the $(p,q)$ signature, i.e. $A = \text{sign} (\Lambda)$
    \item 
    Compute the the embedding into $\mathbb{R}^{p,q}$ by finding $VAV^T$.
    \item 
    For each point $x$ in the embedding, split it into $x^{(p)}$ and $x^{(q)}$. Use standard JL transform to project into $\mathbb{R}^{p'}$ and $\mathbb{R}^{q'}$ respectively where $p'$ and $q'$ are the specified target dimension.
    \item 
    Return the projected points $(x^{(p')}, x^{(q')})$ along with their $(p', q')$ signature.
\end{enumerate}
\end{tbox}

\begin{tbox}
    \textbf{Power Distance JL Transform}
    \begin{enumerate}
        \item  
        Find the orthogonal decomposition of the dissimilarity matrix $O\Lambda O^T$
        \item 
        Compute $e_n$, the smallest eigenvalue of $\Gram(D)$. Set $r= \sqrt{|e_n|}/2$.
        \item 
        Add $4r^2(J - I)$ to $D$ to obtain the new Euclidean distance matrix, $E$.
        \item 
        Recover the Euclidean coordinates $X'$ such that $E_{ij}=\|x'_i-x'_j\|^2$. Perform standard JL transform on $X'$ to target dimension $m$.
        \item 
        Return the projected points along with their radii $r$.
    \end{enumerate}
\end{tbox}

\textbf{Computational efficiency.} Since both methods start with a dissimilarity matrix we necessarily need to recover the `coordinates' first. This can be done by running the singular value decomposition (SVD) on the Gram matrix in $O(n^3)$ time. 
There are several possible ways to speed up this step in practice – by using landmark MDS for example. Another approach, motivated by the generalized power distance formulation, is to consider appending a proper term $4r^2$ to the input matrix $D$ and later use the power distance with the recovered coordinates. If it happens that the coordinates are to be learned from a representation learning module, we can skip this part and directly apply dimension reduction. When 
applying standard JL transform we use random projection of Gaussian vectors. 
One can use any JL transform for example~\cite{Achlioptas2003-pg,Ailon2009-of,Dasgupta10.1145/1806689.1806737,yin2020extremely,Kane2014-rn} for this purpose.

\section{Experiments}
\label{sec:exp}

In this section, we present experimental results of the proposed JL transforms in non-Euclidean settings with only a dissimilarity matrix as input. First, we validate our theoretical results by showing the approximation error on datasets that are highly non-Euclidean. Next, we evaluate the algorithms on real-world datasets, with the classical JL transform as a baseline. Finally, we test the proposed JL transforms on a downstream task, $k$-means clustering, to show its potential wide applications.

\textbf{Datasets.}
We use two \textbf{synthetic datasets} that are made non-Euclidean: Random-simplex and Euclidean-ball. At a high level, for the Random-simplex, given a dataset of size $n$, each point is constructed such that its first $n-1$ coordinates form a simplex, while the final coordinate dominates the pairwise distances. This design induces a large negative eigenvalue in the Gram matrix. The Euclidean-ball dataset, inspired by Delft’s balls~\cite{Duin2010-wh}, consists of $n$ balls with varying radii. The distance between two balls is defined as the minimal distance between any two points on their surfaces, resulting in dissimilarities that violate the triangle inequality.
For \textbf{real-world data}, We consider three categories: genomics, image and graph data. The genomics data includes three cancer-related datasets from the Curated Microarray Database (CuMiDa)~\cite{feltes2019cumida}. Following the practice in prior work~\cite{deng24NeucMDS}, we obtain dissimilarities with entropic affinity. We also test two celebrated image datasets: MNIST and CIFAR-10, each with 1000 images randomly sampled. We use the measures mentioned in~\cite{how_MDS_wrong_NEURIPS2021} to calculate the dissimilarities. The graph datasets are selected from the SNAP project~\cite{snapnets}. The pairwise distances are shortest path distances. The basic statistics of all datasets are shown in \Cref{tab:dataset}. We defer more details in Appendix~\ref{appendix:experiments}.

\begin{table}[h]
\centering
\renewcommand{\arraystretch}{1.3}
\begin{adjustbox}{width=\linewidth}
\begin{tabular}{c|cc|ccc|cc|ccc}
\thickhline
 Dataset & Simplex  &  Ball & Brain & Breast  & Renal  & MNIST & CIFAR10 & Email & Facebook & Mooc  \\
\hline
 Size & 1000 & 1000 & 130 & 151 & 143 & 1000 & 1000 & 986 & 4039 & 7047 \\
 \# $\{\lambda < 0\}$ & 900 & 887 & 53 & 59 & 57 & 454 & 399 & 465 & 1566 & 268\\
 Metric & \ding{55} & \ding{55} & \ding{55}  & \ding{55} & \ding{55} & \ding{51} & \ding{51} & \ding{51} & \ding{51} & \ding{51}\\
 \thickhline
\end{tabular}
\end{adjustbox}
\vspace{1mm}
\caption{Non-Euclidean/Non-metric Datasets used in experiments}
\label{tab:dataset}
\end{table}


\subsection{Validation of Theoretical Results}

\begin{figure}[htbp]
    \centering
    \begin{minipage}[b]{0.49\linewidth}
        \centering
        \includegraphics[width=\linewidth]{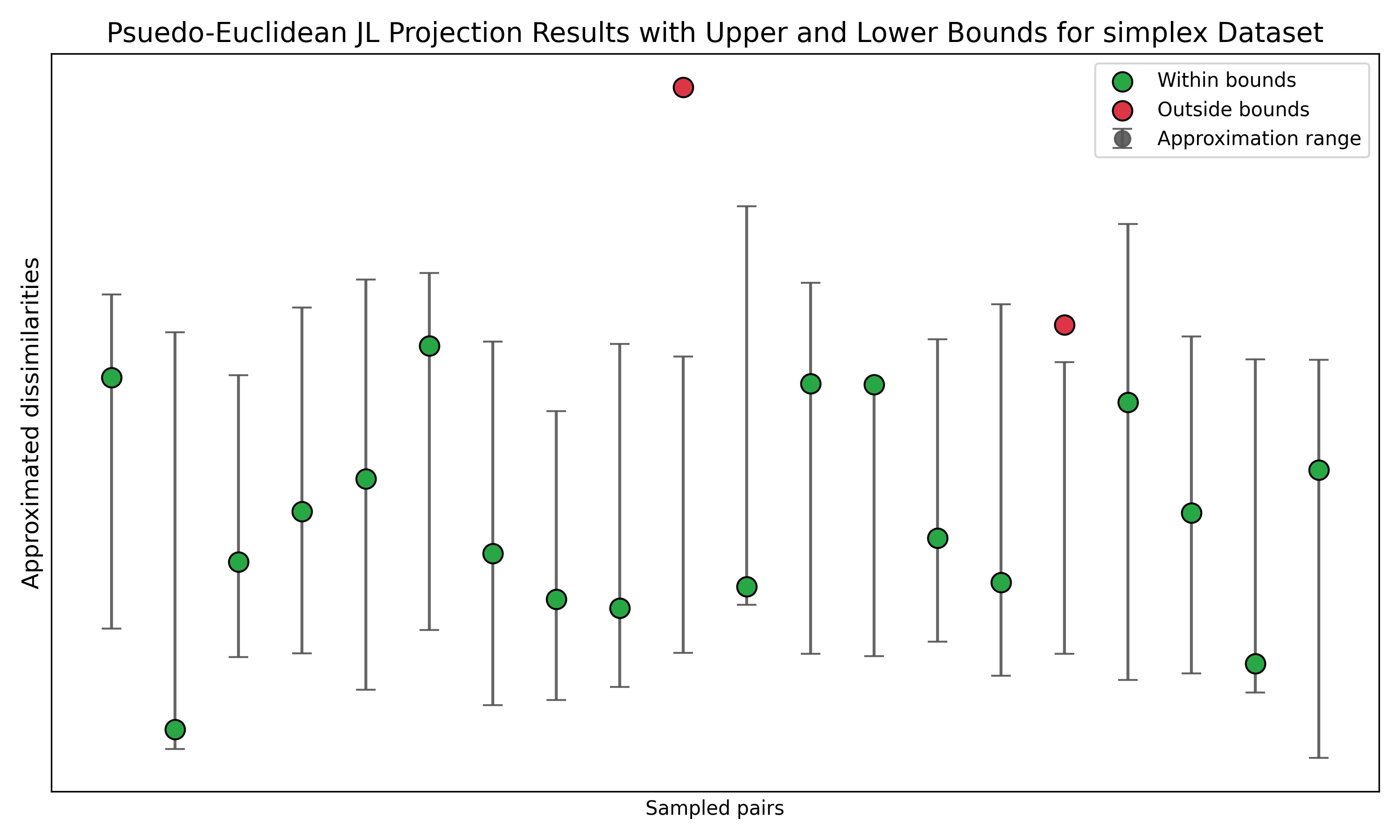} 
    \end{minipage}
    \hfill
    \begin{minipage}[b]{0.49\linewidth}
        \centering
        \includegraphics[width=\linewidth]{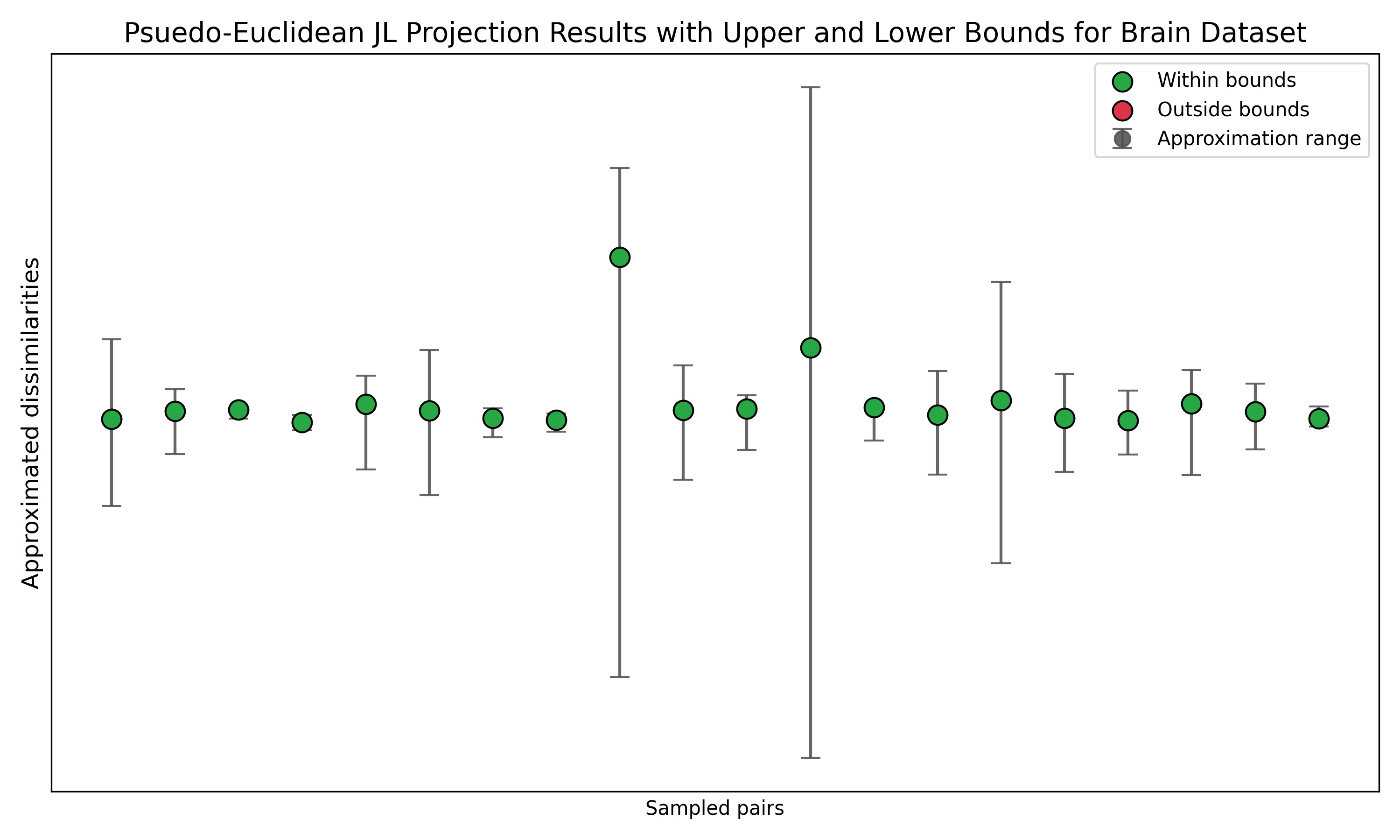}  
    \end{minipage}
    \caption{The left is for Simplex dataset and right for the Brain dataset. Green points indicate the approximation ratio is within the range and red the opposite. The error bars match the upper and lower bounds given by $(1\pm \eps\cdot C_{{ij}})$. We sample 20 pairs of dissimilarities for presentation.}
    \label{fig:jl-pq-synthetic-thm}
\end{figure}

\textbf{Pseudo-Euclidean JL Transform.} We start with demonstrating the multiplicative factor of pseudo-Euclidean projection indeed falls into the range of $(1\pm \eps\cdot C_{{ij}})$, as a corroboration of \Cref{thm:jl-pq}. Throughout, we set $\varepsilon=0.5$ and the constant in the target dimension $O(\log n/\eps^2))$ as 2. \Cref{fig:jl-pq-synthetic-thm} verifies the approximation ratio on the Simplex and Brain datasets. We observe that most pairwise dissimilarities behave as claimed. For the Brain dataset, even the bar range is very small we still have all green points. There are minor exceptions indicated by the red points in the Simplex dataset. However, first observe that they are still very close to the error bars. Further, we did not optimize the constant factor in $O(\log n/\eps^2))$. In fact, if we double the target dimension (from 80 to 160), the percentage of violations drops significantly from 12.01\% to 4.62\% for the Simplex dataset. We defer the plots of other datasets to \Cref{appendix:experiments}.

\begin{figure}[htbp]
    \centering
    \begin{minipage}[b]{0.49\linewidth}
        \centering
        \includegraphics[width=\linewidth]{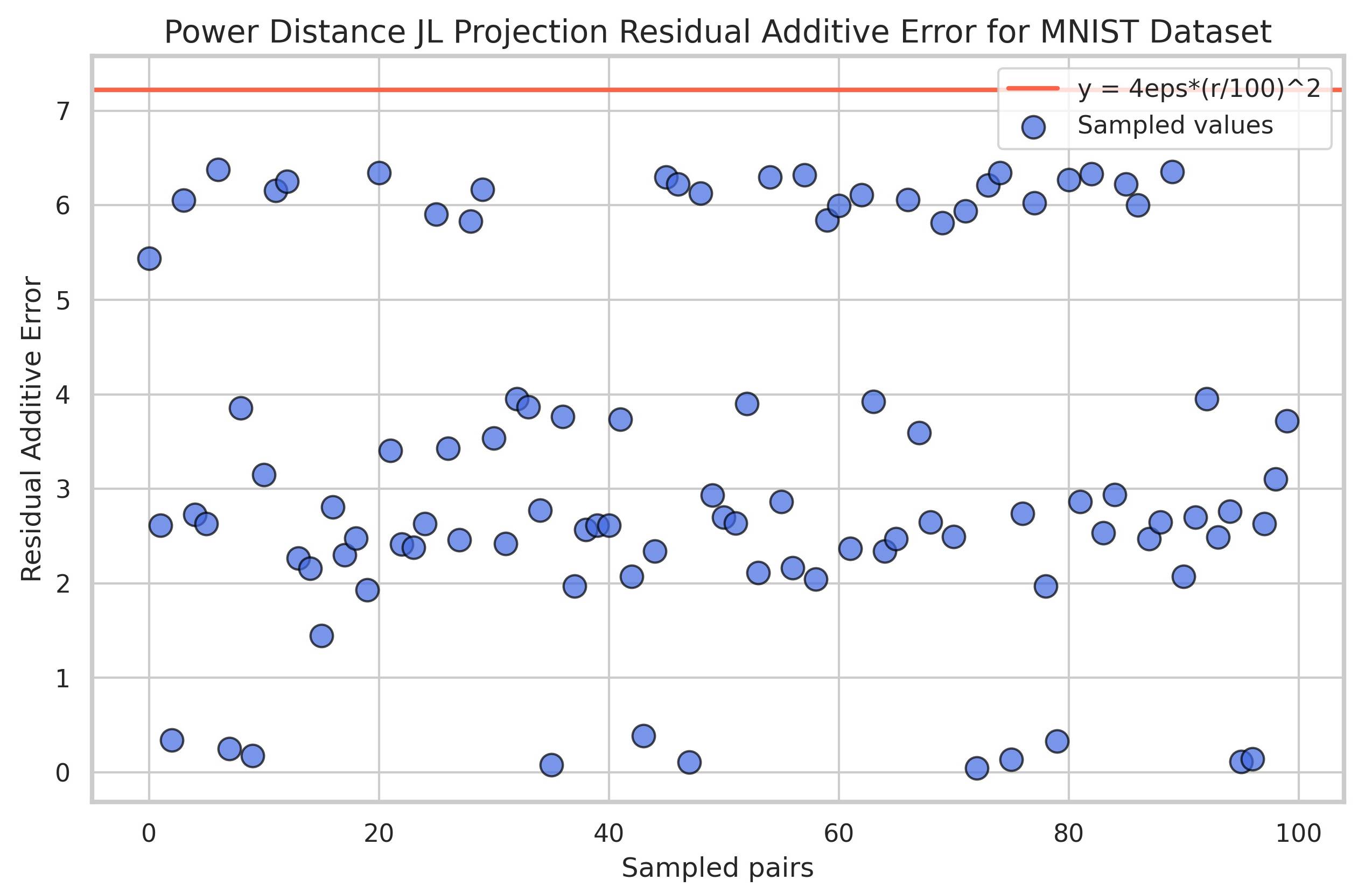} 
    \end{minipage}
    \hfill
    \begin{minipage}[b]{0.49\linewidth}
        \centering
        \includegraphics[width=\linewidth]{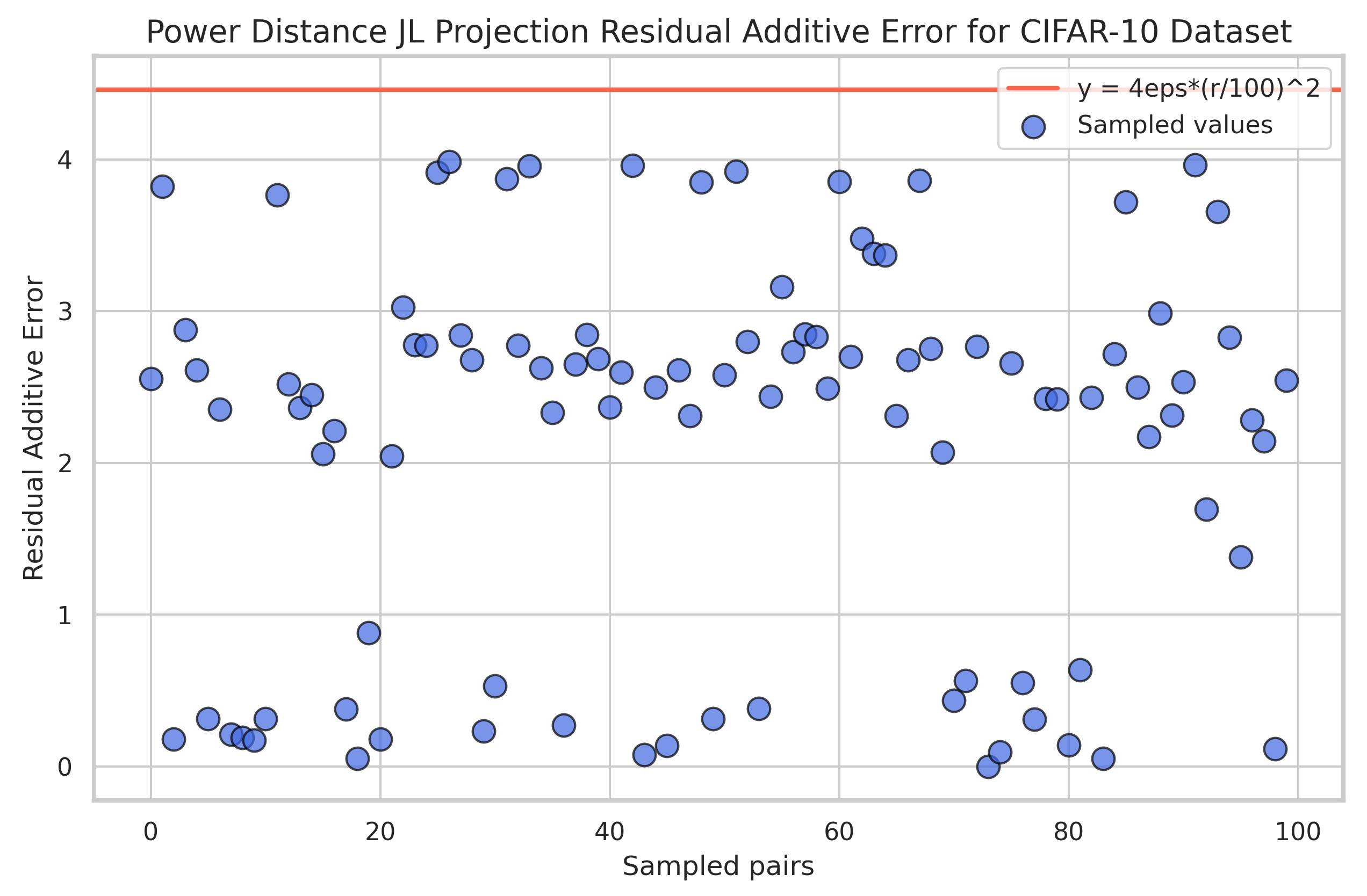}  
    \end{minipage}
    \caption{The left is for MNIST and right for CIFAR-10 dataset. We sampled 100 pairs of dissimilarities and adopt $r/100$ for presentation. Using the original $r$ makes the red line at roughly $y = 40000$, which is very loose.}
    \label{fig:jl-power-thm}
\end{figure}

\textbf{Power Distance JL Transform.} To validate \Cref{thm:JL-power}, we show the residual additive error of the power distance JL transform. By 'residual' we refer to the absolute difference between the dissimilarities given by the JL transform and the $(1\pm \eps)$ approximation of the true dissimilarities. This term should be smaller than $4\eps r^2$. As shown in \Cref{fig:jl-power-thm}, all sampled points appear below the red line, which is taken as $4\eps(r/100)^2$, a much tighter upper bound. Recall that a larger $r$ indicates more deviation from Euclidean geometry, the fact that the residual error is small shows the effectiveness of our JL transform. We defer the plots of other datasets to \Cref{appendix:experiments}.

\subsection{Performance of JL Transforms}
\textbf{Performance on Relative Error.} We compare two proposed JL transforms with the classical JL transform on all datasets and report the relative error, which is defined as the maximum among $\frac{|D_{ij} - \hat{D}_{ij}|}{D_{ij}}$ for all $(i,j)$ pairs. $\hat{D}_{ij}$ is the dissimilarity matrix obtained from any JL transform. It is a suitable metric because all three algorithms have different theoretical guarantees, but the end goal is always preserving pairwise dissimilarities. In \Cref{tab:relative-error-all} we show the worst-case (defined above) average and median relative error. A smaller value indicates better performance.

We observe JL-PE performs the best for genomics data and JL-power performs the best for the rest, on both metrics. The significant improvement implies when the JL transform matches with geometry, we can expect really good results. The image datasets report \textit{inf} for JL because different images are projected to the same point. In a few cases Jl-PE has slightly worse relative error than JL, but note that this might be the outcome of really large factor $C$, which the performance of JL-PE is based on.

\begin{table}[h]
\centering
\renewcommand{\arraystretch}{1.3}
\begin{adjustbox}{width=\linewidth}
\begin{tabular}{c|cc|ccc|cc|ccc}
\thickhline
 Method & Simplex  &  Ball & Brain & Breast & Renal  & MNIST & CIFAR10 & Email & Facebook & MOOC  \\
\hline
 JL Max & 6.47e5 & 5.97e5 & 1.02e12  & 7.09e10 & 3.42e12 & inf & inf & 3.85e4 & 1.18e6 & 2.94e5     \\
 JL-PE Max & 1.11e6 & 5.31e5 &  \textbf{8,21e4} & \textbf{262.20} & \textbf{2.37e4} & 8.47e5 & 2.24e6 & 3.24e6 & 2.51e7 & 1.35e7    \\
 JL-Power Max & \textbf{109.18} & \textbf{12.102} &  3.11e7 & 1.92e6 & 1.17e8 & \textbf{85.76} & \textbf{55.74} & \textbf{781.45} & \textbf{82.74} & \textbf{24.22}     \\
 \hline
 JL Ave & 442.65 & 43.78 &  1.11e9 & 9.86e7 & 1.50e9 & inf & inf & 15.64 & 12.62 &  39.83    \\
 JL-PE Ave  & 47.59 & 23.71 & \textbf{8.16}  & \textbf{1.15} & \textbf{6.60} & 15.93 & 18.24 & 13.79 & 52.52 & 63.44   \\
 JL-Power Ave & \textbf{13.52} & \textbf{1.002} &  3.73e4 & 3.12e3 & 5.57e4 & \textbf{1.47} & \textbf{1.61} & \textbf{1.60} & \textbf{1.32} & \textbf{2.04} \\
  \hline
 JL Median & 6.36 & 6.68 & 2.69e12 & 1.95e11 & 1.18e12 & 1.78 & 2.29 & 7.00 & 5.08 & 6.13   \\
 JL-PE Median  & \textbf{2.57} & 2.78 & \textbf{1284.06} & \textbf{174.74} & \textbf{3160.58} & 2.11 & 2.04 & 2.06 & 5.83 & 6.03    \\
 JL-Power Median & 12.79 & \textbf{1.00} &  1.01e7 & 8.20e6 & 7.57e7 & \textbf{1.30} & \textbf{1.35} & \textbf{1.35} & \textbf{1.16} & \textbf{1.99} \\
 \thickhline
\end{tabular}
\end{adjustbox}
\vspace{1mm}
\caption{Max and Average Relative Error of All Datasets on Three JL Transforms.}
\label{tab:relative-error-all}
\end{table}

\textbf{Performance on Downstream $k$-Means Clustering.} We now evaluate if the propsoed JL transforms consistently perform well on downstream tasks. The setup is straightforward: we first apply JL transform to reduce the dimension of the original data, then we compare $k$-Means clustering cost obtained from the processed data with the original optimal clustering cost.

\begin{table}[h!]
\centering
\small
\begin{tabular}{cccccc}
\thickhline
\textbf{Dataset} & \textbf{\#Clusters (k)} & \textbf{Original} & \textbf{JL} & \textbf{JL-pq} & \textbf{JL-power} \\
\hline
Brain & 5 & $3.92 \times 10^{-3}$ & 728.00 & \textbf{3.13} & 33.26 \\
Breast & 4 & 0.03 & 827.69 & 28.93 & \textbf{12.53} \\
Email & 42 & 5614 & 22634 & 21244 & \textbf{20764} \\
Facebook & 15 & 20499 & 284142 & \textbf{91644} & 303203 \\
\thickhline
\end{tabular}
\vspace{1mm}
\caption{Performance of Three JL Transforms with $k$-Means Clustering.}
\label{tab:results-clustering}
\end{table}

\Cref{tab:results-clustering} demonstrates that both of the proposed JL transforms outperforms the classical JL, with only one exception on Facebook where JL-power is slightly worse than JL. This observation shows better effectiveness of the proposed JL transforms in downstream tasks. However, similar as above, the relative performance varies between JL-pq and JL-power.  This raises a key direction for future research: formalize the conditions under which a specific JL transform is most suitable.



\section{Discussions}
The Pseudo-Euclidean space we study shares conceptual foundations with recent advances in similarity learning. Notably, ~\cite{Liu2019-ya} proposed neural similarity learning (NSL) which uses bilinear matrices to generalize inner product similarity in convolutional neural networks through both static and dynamic approaches. 
The bilinear form structures also appear in modern neural architectures ~\cite{vaswani2017attention}. In the scaled dot-product attention mechanism used in transformers, the attention score between two tokens is computed as a bilinear form:
$\text{score}(q, k) = q^\top W k $, where $q$ and $k$ are query and key vectors and $W$ is a learned weight matrix. This mechanism effectively learns a similarity function over the token space, which aligns with our use of bilinear forms to represent pairwise dissimilarities in non-Euclidean geometry. 

\textbf{Future Direction.} A promising future direction is to integrate the mathematical formulation in this paper using both the pseudo-Euclidean representation and the generalized power distance representation with current machine learning pipelines for representation and similarity learning. 
Recently there has been increasing interest in representation learning in non-Euclidean spaces, e.g., learning modules in hyperbolic ~\cite{nickel17poincare,gulcehre2018hyperbolic,Ganea2018hyperbolic,tifrea2018poincare,De-Sa2018-cc,Chami2020-ug,Chami2019-ie}, spherical or mixed-curvature product manifolds (Cartesian products of hyperbolic, hyperspherical and Euclidean manifolds)~\cite{GuSGR19,pmlr-v130-zhang21j,mcneela2025product,chlenski2025manifypythonlibrarylearning} or general manifolds~\cite{Aalto19metric}. Many of the prior work assume (piecewise) constant curvature in the embedded space and our setting includes these spaces and beyond. One limitation of current work is we are not able to build a theoretical connection between two proposed representations, or conditions that one outperforms the other. Further, we are not aware of any lower bound result complementing the fine-grained JL guarantee from the pseudo-Euclidean representation.

\textbf{Conclusion.} We explore Johnson-Lindenstrauss Transform in the non-Euclidean setting with dissimilarity matrix as the input. Via two different approaches, pseudo-Euclidean geometry and generalized power distance, we show two theoretical results with similar flavor to the classical JL lemma. The first gives a fine-grained error analysis and the second has an extra additive error term, both having parameters indicating the deviation from Euclidean geometry. The experiment results corroborate with our theoretical results, and demonstrate the superior performance of two proposed approaches for JL-type dimension reduction with non-Euclidean data.

\section{Acknowledgements}
The authors would like to acknowledge funding from IIS-22298766, DMS-2220271, DMS-2311064, DMS- 2501286,  CRCNS-2207440, CCF-2118953 and CNS-2515159. The authors would like to thank Hongbin Sun for helpful discussions.

\newpage
\bibliographystyle{alpha} 
\bibliography{pqspace,references,references-MDS,dataset}

\newpage
\appendix

\section{Relation of Generalized Power Distance to Silhouette Coefficient}\label{appendix:silhouette}

The silhouette coefficient~\cite{Silhouettes_Rousseeuw1987} is a measure of how similar an object is to its own cluster compared to other clusters. The silhouette coefficient ranges from -1 to 1, where a high value indicates that the object is well matched to its own cluster and poorly matched to neighboring clusters. If most objects have a high value, then the clustering configuration is appropriate. If many points have a low or negative value, then the clustering configuration may have too many or too few clusters. 

For simplicity, let us consider two clusters of finite points $X, Y\subseteq \mathbb{R}^m$ with $|X|=|Y|=n$. The classical silhouette coefficient (in two clusters) can be defined as follows: 
For each point $x\in X$, define ``\disin'' $a(x)$ as the average distance between $x$ and all other points in the same cluster, and ``\disout'' $b(x)$ as the average distance between $x$ and all points in the other cluster $Y$. That is:  
\begin{align}
    a(x) &= \frac{1}{n - 1}\sum_{x' \in X,\, x' \neq x} d(x, x')\\
    b(x) &= \frac{1}{n}\sum_{y \in Y} d(x, y)
\end{align}
Then, the silhouette coefficient for a point $x\in X$ is defined as:      
\[\bar{s}(x) = \frac{b(x) - a(x)}{\normalizer(a(x), b(x))},\] 
and the silhouette coefficient for the cluster $X$ is defined as the average of the silhouette coefficients for all points in the cluster:
\[\bar{s}(X) = \frac{1}{n}\sum_{x \in X} \bar{s}(x).\]
Here the denominator $\normalizer$ is a normalizing function which is usually chosen to be  $\normalizer(a(x), b(x)) = \max\{a(x), b(x)\}$ for the classical silhouette coefficient to make sure the range lies in $[-1, +1]$. The silhouette coefficient can be used to measure the quality of a cluster. For the overall clustering, the average silhouette coefficient scores over all clusters can be used to measure the quality of the whole clustering results~\cite{KaufmanR90ClusterAnalysis}.


Now we are interested in using a similar idea to study the dissimilarity (divergence) between two probability distributions, say two Gaussian distributions $\gX=\gN(\mu_x, \sigma_x), \gY=\gN(\mu_y, \sigma_y)$. 
We can naturally view the two distributions as two clusters. Here we are interested in the ``silhouette coefficient score'' of the whole clustering.

To better analyze the ``silhouette coefficient score'' between two Gaussian distributions, we consider an unnormalized version of the silhouette coefficient. 
For any $x\in \gX$, let 
\begin{align*}
    a(x) &= \sE_{x'\sim \gX} [\|x- x'\|^2],\\
    b(x) &= \sE_{y\sim \gY} [\|x-y\|^2],\\
    s(\gX) &=\sE_{x\sim \gX} [b(x) - c \cdot a(x)],
\end{align*}
where $c \geq 1$ is a constant parameter used as a trade-off balance between {\disin} and {\disout}.
Now we can define an (unnormalized) silhouette score between two Gaussian distributions $\gX$ and $\gY$ as:
\begin{align*}
    SC(\gX, \gY) &\triangleq \frac{1}{2}[s(\gX) + s(\gY)]\\
    &=\frac{1}{2}[\sE_{x\sim \gX} [b(x) - c \cdot a(x)] + \sE_{y\sim \gY} [b(y) - c \cdot a(y)]]\\
    &=\sE_{x,y\sim \gX\times \gY} \big[\|x-y\|^2\big] - \frac{c}{2} \left( \sE_{x,x'\sim \gX}[\|x-x'\|^2] + \sE_{y,y'\sim \gY}[\|y-y'\|^2]  \right)
\end{align*}
Based on properties of Gaussian distributions, we could calculate:
\begin{equation*}
    SC(\gX, \gY) = \|\mu_x- \mu_y\|^2 - (c-1) (\sigma_x^2 + \sigma_y^2).
\end{equation*}
In general, we could consider a silhouette score between two Gaussian distributions composed with two parts balanced with each other:
\begin{enumerate}
    \item Some distance $\dis_\sP(\gX, \gY)$ measures the similarity between the two distributions.
    \item Some negative terms $\Delta$ represent the variances of the two distributions themselves.  
\end{enumerate}
Here we consider the following construction of a silhouette score between two Gaussian distributions we are interested in: Let $\Delta[\gX]=\sE_{x,x'\sim \gX}[\|x-x'\|^2]$.
\begin{align}
    SC_W(\gX, \gY) &\triangleq \dis_W^2(\gX, \gY) - (\Delta[\gX]+ \Delta[\gY])    \\
    &= \|\mu_x-\mu_y\|^2 + (\sigma_x - \sigma_y)^2 - 2(\sigma_x^2 + \sigma_y^2) \\
    &= \|\mu_x- \mu_y\|^2- (\sigma_x + \sigma_y)^2 
\end{align}
where $\dis^2_W$ is the squared $2$-Wasserstein distance. This construction is interesting because it is consistent with our construction of generalized power distance, which brings up a potential statistical interpretation.

\paragraph{Normalized Silhouette Score}

One could also define a normalized version of the Silhouette Score. In general, it could be defined as:
\begin{equation}
    \overline{SC}(\gX, \gY) \triangleq \frac{\dis_\sP(\gX, \gY) - C\cdot (\Delta[\gX ] + \Delta[\gY ] )}{\normalizer(\dis_\sP(\gX, \gY),  \;\; C\cdot (\Delta[\gX ] + \Delta[\gY ])) }
\end{equation}
with some normalizer function $\normalizer()$.

For example, we could define the normalized version $SC_W$ as:
\begin{equation}
    \overline{SC}_W(\gX, \gY) \triangleq \frac{\|\mu_x- \mu_y\|^2  - (\sigma_x + \sigma_y)^2}{\|\mu_x- \mu_y\|^2  + (\sigma_x + \sigma_y)^2}
\end{equation}
The range of $\overline{SC}_W$ is $[-1, 1]$. Intuitively, it can be used to measure how well two Gaussian distributions are separated. 
When these two Gaussian distributions have very close centers or very large variances, the normalized silhouette score is close to $-1$, which means sampling from these two distributions has a low chance of being well separated. It becomes $-1$ when they are identical $\gX=\gY$. When these two distributions are far away or variances are very small, the normalized silhouette score is close to $1$, which means sampling from these two distributions enjoys a high chance of being well separated. It becomes $+1$ when these two distributions are delta distributions on distinct center points. Zero represents borderline cases.
We set it to be $-1$ for $\mu_x=\mu_y$ and $\sigma_x=\sigma_y=0$. 



\section{Missing Proofs in \Cref{sec:jl-pq} and \Cref{sec:jl-power}} \label{sec:proofs}

\subsection{Proofs in \Cref{sec:jl-pq}}

\begin{proof}[Proof of Lemma \ref{lem:jl-pq}]
We apply the standard Johnson-Lindenstrauss lemma to the collection of points $\{x_i^{(p)}\}$ and $\{x_i^{(q)}\}$ separately.  By~\Cref{eq:JL} there are two random projection $f_p:\mathbb{R}^p \rightarrow \mathbb{R}^{p'}$ and $f_q:\mathbb{R}^q \rightarrow \mathbb{R}^{q'}$ with $p', q' = O(\log n/\eps^2)$ such that for any $i\neq j$,
\begin{align*}
(1-\eps)\enorm{x_i^{(p)}-x_j^{(p)}}^2\leq \enorm{f_p(x_i^{(p)})-f_p(x_j^{(p)})}^2\leq (1+\eps)\enorm{x_i^{(p)}-x_j^{(p)}}^2\\
(1-\eps)\enorm{x_i^{(q)}-x_j^{(q)}}^2\leq \enorm{f_q(x_i^{(q)})-f_q(x_j^{(q)})}^2\leq (1+\eps)\enorm{x_i^{(q)}-x_j^{(q)}}^2
\end{align*}
Now we can bound the $(p',q')$-norm of the vector under projection $f$ which produces $f(x_i)=(f_p(x_i), f_q(x_i))$ of dimension $p'+q'$.
\begin{align*}
    \|f(x_i)-f(x_j)\|^2_{p', q'}&=\enorm{f_p(x_i)}^2-\enorm{f_q(x_i)}^2\\
    &\leq (1+\eps)\enorm{x_i^{(p)}-x_j^{(p)}}^2-(1-\eps)\enorm{x_i^{(q)}-x_j^{(q)}}^2\\
    &=\pqnorm{x_i-x_j}^2+\eps \enorm{x_i-x_j}^2
\end{align*}
The other direction is similar. 
\end{proof}

\begin{proof}[Proof of Lemma~\ref{lem:highp-lowq}]
    Consider the variables $\enorm{v}^2/(p+q)$ and $\pqnorm{v}^2/(p+q)$. Let $\mu = \mathbb{E}(X^2)$ where $X$ is a random variable with distribution $f$. Taking $p + q \to \infty$ and using Central Limit Theorem implies that $\enorm{v}^2/(p+q)$ and $\pqnorm{v}^2/(p+q)$ will converge to their means which are $\mu$ and $\frac{p-q}{p+q}\mu$ respectively. Plugging in $q < \frac{C-1}{C+1} p$ and $p < \frac{C-1}{C+1} q $ give us what we want.
\end{proof}

\begin{proof}[Proof of Theorem \ref{thm:random-pts-pq}]
We may first write the uniform distribution over the sphere as 
$$Y = (\frac{Y_1}{\sqrt{\sum_i^{p+q}Y_i^2}}, \frac{Y_2}{\sqrt{\sum_i^{p+q}Y_i^2}}, ... \frac{Y_{p+q}}{\sqrt{\sum_i^{p+q}Y_i^2}})$$
where each $Y_i = N(0, 1)$. Then the distribution of $\frac{\enorm{u - v}}{\pqnorm{u-v}}$ will be :
\[\frac{\sum_{i=1}^{p+q} Y_i^2}{\sum_{i=1}^{p} Y_i^2-\sum_{i=p+1}^{p+q} Y_i^2}\]
Now $\sum_{i=1}^{p+q} Y_i^2$ is just a chi-squared random variable with $p+q$ degrees of freedom. A common known concentration inequality is:
\[\mathbb{P}(\chi^2 \in (1\pm\delta)(p+q)) \geq 1 - 2e^{-\frac{p+q}{2}(\frac{1}{2}\delta^2-\frac{1}{3}\delta^3)}\]
On the other hand, $\sum_{i=1}^{p} Y_i^2-\sum_{i=p+1}^{p+q} Y_i^2$ has mean $p - q$ and we have the following concentration inequality:
\[\mathbb{P}(\chi^2 \in (p-q)\pm 2\delta(p+q)) \geq 1 - 2e^{-\frac{p+q}{2}(\frac{1}{2}\delta^2-\frac{1}{3}\delta^3)}\]
since $\sum_{i=1}^{p+q} Y_i^2 \in (1\pm\delta)(p+q)$ implies $\sum_{i=1}^{p} Y_i^2-\sum_{i=p+1}^{p+q} Y_i^2 \in (p-q)\pm 2\delta(p+q)$. The greatest error ratio clearly is bounded by $q = \frac{C-1}{C+1}p$ and at:
\[\frac{(1+\delta)(p+q)}{(p-q) - 2\delta(p+q)}\frac{p-q}{p+q} - 1 = \frac{2pC(1+\delta)/(C+1)}{2p/(C+1) - 4\delta pC/(C+1)}\frac{1}{C}-1 = \frac{\delta + 2\delta C}{1-2\delta C}\]
\[\mathbb{P}(\frac{\sum_{i=1}^{p+q} Y_i^2}{\sum_{i=1}^{p} Y_i^2-\sum_{i=p+1}^{p+q} Y_i^2} \in (1\pm \frac{\delta + 2\delta C}{1-2\delta C})\frac{p+q}{p-q}) \geq 1 - 2e^{-\frac{p+q}{2}(\frac{1}{2}\delta^2-\frac{1}{3}\delta^3)}\]
Now let $\delta = \frac{\epsilon/2}{4C + 2 + \epsilon C}$:
\[\mathbb{P}(\frac{\sum_{i=1}^{p+q} Y_i^2}{\sum_{i=1}^{p} Y_i^2-\sum_{i=p+1}^{p+q} Y_i^2} \in (1\pm \frac{\epsilon}{2})\frac{p+q}{p-q}) \geq 1 - 2e^{-O((p+q)\epsilon^2/C^2)} \geq 1 - O(n^{-2})\]
The last inequality comes from the fact that our target dimension is $O(C^2\log(n)/\epsilon^2)$, so we can make the assumption $p+q > O(C^2\log(n)/\epsilon^2)$. From a union bound, then we have with high probability, $\frac{\enorm{u - v}}{\pqnorm{u-v}} \leq C(1 + \frac{\epsilon}{2})$
Then, we can use \Cref{lem:jl-pq} with input $\epsilon' = \epsilon / (2C)$ which gives us the error bound:
\begin{align}
    \|f(x_i)-f(x_j)\|^2_{p', q'} & \leq \pqnorm{x_i-x_j}^2+\frac{\epsilon}{2C} \enorm{x_i-x_j}^2\\
                                 & \leq \pqnorm{x_i-x_j}^2+\frac{\epsilon}{2C} C(1 + \frac{\epsilon}{2})\pqnorm{x_i-x_j}^2\\
                                 & = \pqnorm{x_i-x_j}^2+(\frac{\epsilon}{2} + \frac{\epsilon^2}{2})\pqnorm{x_i-x_j}^2\\
                                 & \leq (1+\epsilon)\pqnorm{x_i-x_j}^2\\
\end{align}
\end{proof}

\subsection{Proofs in \Cref{sec:jl-power}}

\begin{proof}[Proof of Lemma \ref{lem:power-D}]
    We begin by finding the Gram Matrix of $E = D - 4r^2(I-J)$. Recall that the centering matrix $C = I - \frac{J}{n}$. The Gram matrix of $E$ is
    \[-\frac{1}{2}CEC=-\frac{1}{2}C(D - 4r^2(I-J))C = -\frac{1}{2}CDC + 2r^2C(I-J)C = -\frac{1}{2}CDC + 2r^2C.\]
    Here we use the fact that $C(I-J)=C$ and $C^2=C$.
    Further, $C=I - \frac{J}{n}$ is symmetric and positive semi-definite. $C$ has a single eigenvalue of $0$ with $\mathbf{1}_n$ as its eigenvector. We see that $\Gram(D)=-\frac{1}{2}CDC $ and $C$ share that eigenvalue and eigenvector. The rest of the eigenvalues of $C$ are 1. Then, the eigenvalues of $\Gram(E)$ are exactly $2r^2$ added to the eigenvalues of $\Gram(D)$. Thus, it is clear all of its eigenvalues are greater than or equal to zero as long as $2r^2 \geq |e_n|$.
\end{proof}

\begin{proof}[Proof of Lemma \ref{lem:JL-power}]
We apply the standard Johnson-Lindenstrauss Lemma on the centers of the power distance representation $\{p_i\}$ with a random projection $f:\mathbb{R}^d \rightarrow \mathbb{R}^m$. By~\Cref{eq:JL} we have
\begin{align*}
    \Pow((f(p_i), r_i), (f(p_j), r_j))& =\|f(p_i)-f(p_j)\|_E^2-(r_i+r_j)^2\\ 
    &\leq (1+\eps)\|p_i-p_j\|_E^2-(r_i+r_j)^2\\
    & = (1+\eps)\Pow((p_i, r), (p_j, r))+\eps(r_i+r_j)^2.
\end{align*}
The other direction is similar. 
\end{proof}
\section{Additional Experimental Results}\label{appendix:experiments}

\subsection{Details of Data Sets}

\paragraph{SNAP dataset.}
 We tested our methods on a diverse collection of real-world network datasets sourced from the Stanford Network Analysis Project (SNAP)~\cite{snapnets}. These datasets span various domains and exhibit a wide range of structural properties, helping to validate the robustness and generalizability of our findings.

 \begin{itemize}
     \item \textbf{Email-Eu-core (Email)}~\cite{LeskovecTKDD2007Email}: Email exchanges within a European research institution.
     \item \textbf{MOOC-actions (MOOC)}~\cite{KumarKDD2019MOOC}: User actions on a Massive Open Online Course platform.
     \item \textbf{Facebook-ego-networks (Facebook)}~\cite{McAuleyNIPS2012Facebook}: Ego-networks of Facebook users.
 \end{itemize}
 We only used the graph structures and ignored the edge weights and directions. That means all graphs are treated as unweighted and undirected. 

 \paragraph{Synthetic datasets for examples.} We also manually generated two datasets to demonstrate when projecting in Pseudo-Euclidean space is better than projecting with power distances and vice versa. In the first dataset, we had a single very large negative eigenvalue and the rest all positive eigenvalues in the diagonal matrix of the orthogonal decomposition of the distance matrix. In this situation, power distances of the points would have large radii and the error would be large while Pseudo-Euclidean projection would be on a signature of $(p, 1)$, so the error bound would be much smaller. In the second dataset, we had the ratio of positive eigenvalues to negative eigenvalues was a constant fraction, and the negative eigenvalues were all small. In this way, Pseudo-Euclidean projection would have a large error bound while power distance projection would have very small radii and a small error bound.

 \paragraph{Other datasets.} We randomly select 1000 images from MNIST and CIFAR-10. The distance measure between two images are computed by the $k$-neareast neighbor (i.e. Isomap) with $k=10$. The distance measure of genomics dataset is Entopic Affinity, introduced in prior works and commonly used for this datasets. The details of implementation can be found in our Github Repo.

\subsection{Details of Implementation}

All experiments are done with a Macbook Pro with Apple M2 Chip of 16GB memory. The algorithms are fairly efficient and we did not encounter issues on computational resources.

The JL Transforms have only one parameter $\eps$, and we set that to be 0.5 throughout the paper. The constant used in the big O notation of $O(\log n/\eps^2)$ is 2. Another caveat is, our setting does not give access to data coordinates, therefore after obtaining coordinates from the JL transform, we use the original dissimilarity measure to reconstruct $\hat{D}$. If the original measure is inexplicit or unknown, we use Euclidean distance. We only use the data coordinates for the classical JL.

\subsection{More Illustrations}
 The illustrations here extend the experiments to verify \Cref{thm:jl-pq} and \Cref{thm:JL-power}.
\begin{figure}[h]
    \centering
    \begin{subfigure}{0.48\linewidth}
        \centering
        \includegraphics[width=\linewidth]{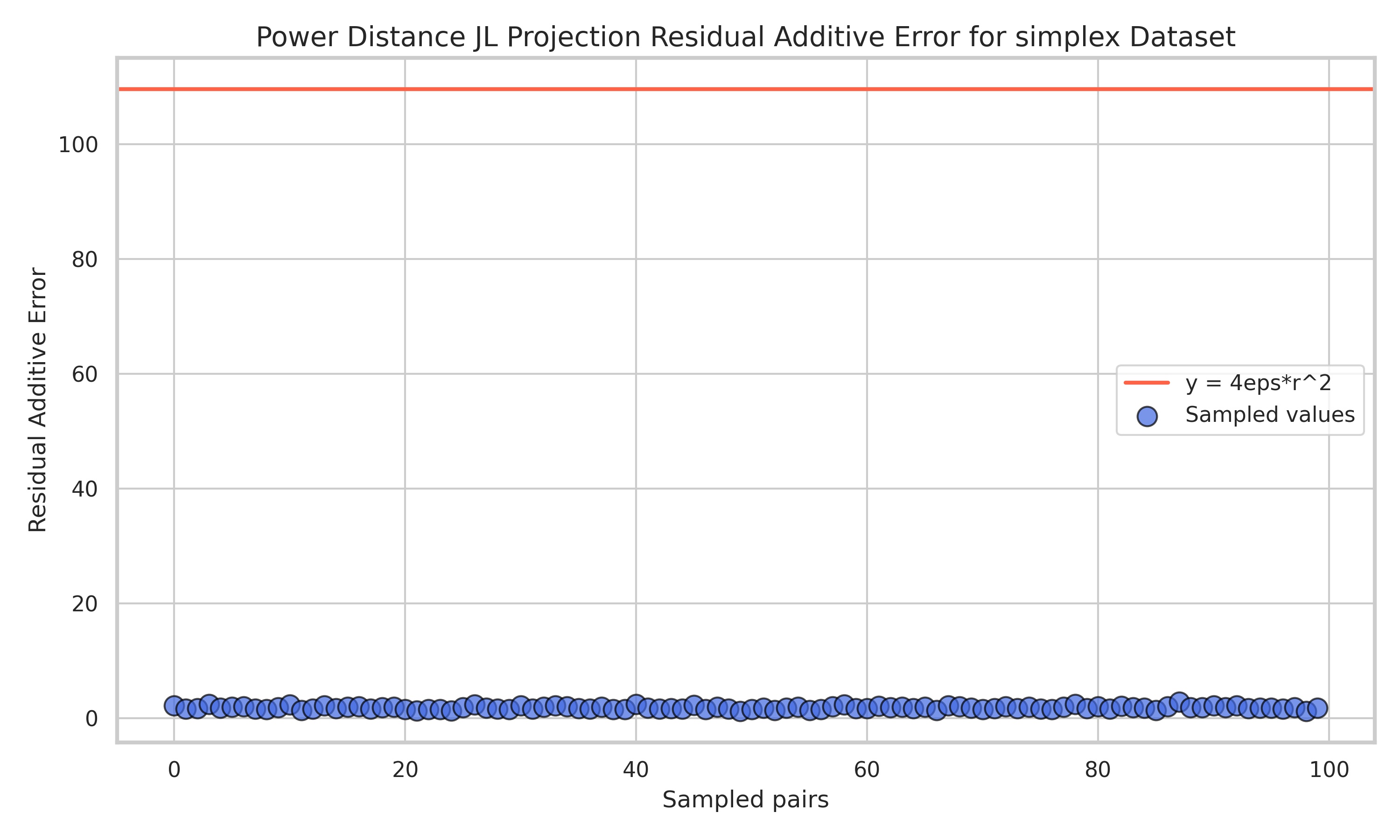}
        \caption{Residual error on simplex}
        \label{fig:sub1}
    \end{subfigure}
    \hfill
    \begin{subfigure}{0.48\linewidth}
        \centering
        \includegraphics[width=\linewidth]{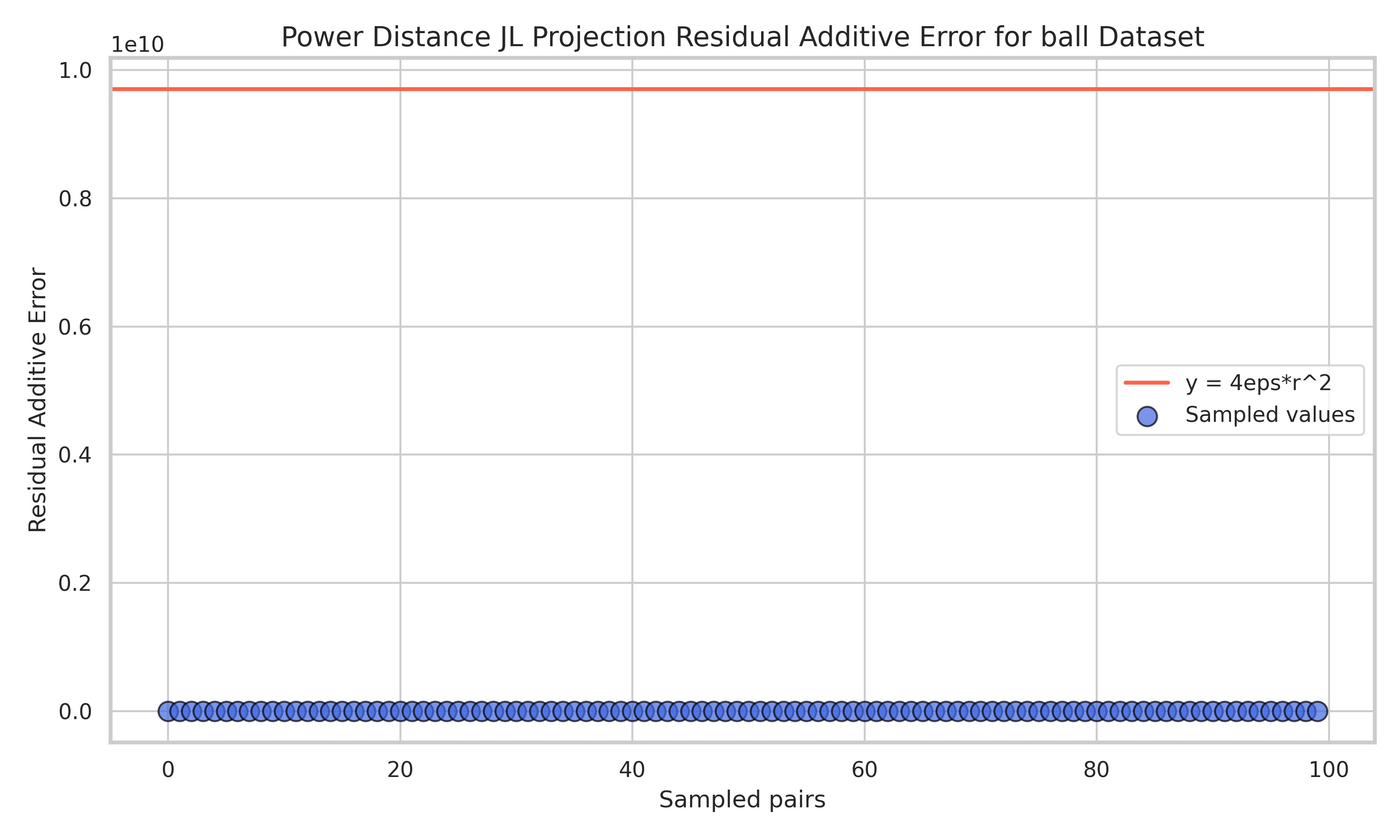}
        \caption{Residual error on ball}
        \label{fig:sub2}
    \end{subfigure}
    
    \vspace{0.5em}
    
    \begin{subfigure}{0.48\linewidth}
        \centering
        \includegraphics[width=\linewidth]{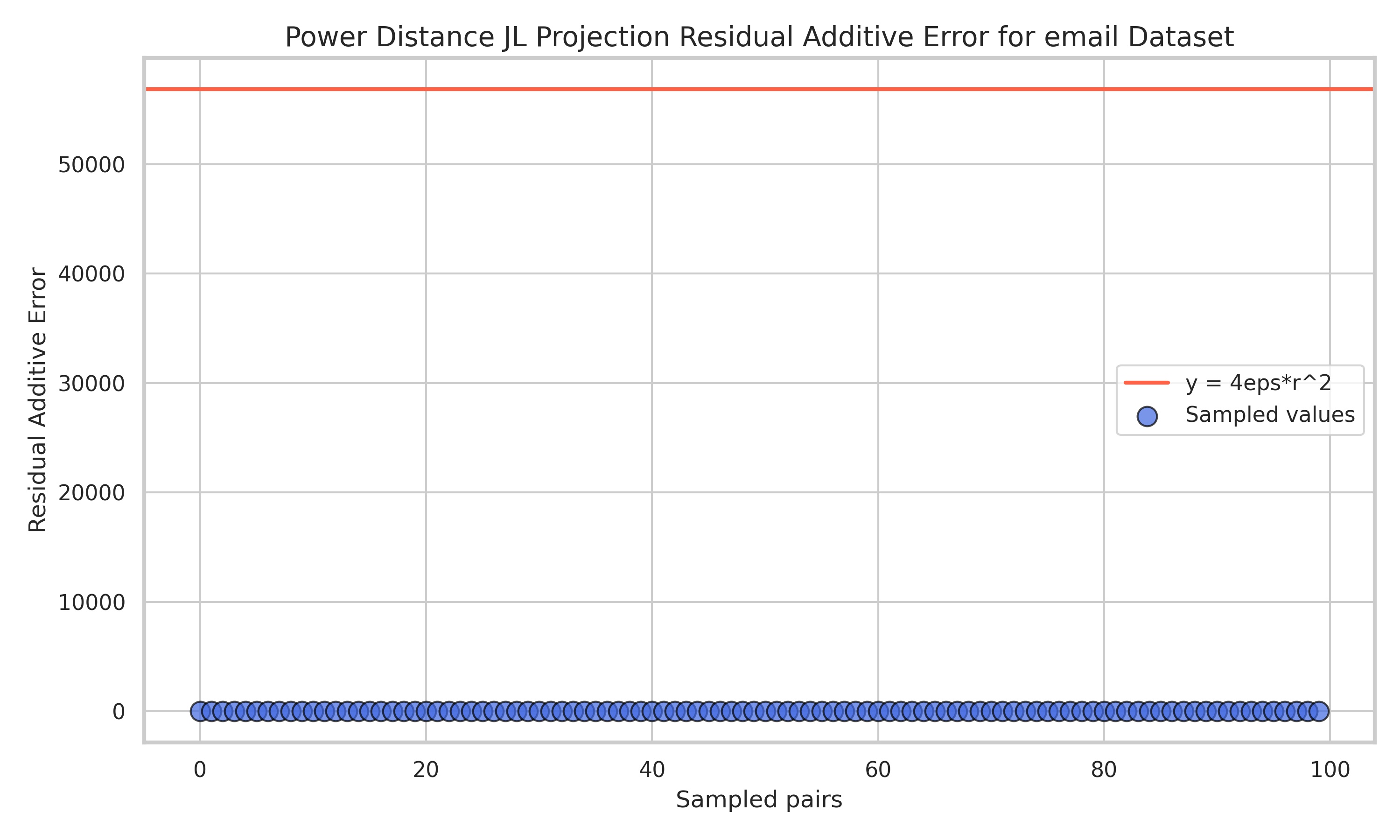}
        \caption{Residual error on email}
        \label{fig:sub3}
    \end{subfigure}
    \hfill
    \begin{subfigure}{0.48\linewidth}
        \centering
        \includegraphics[width=\linewidth]{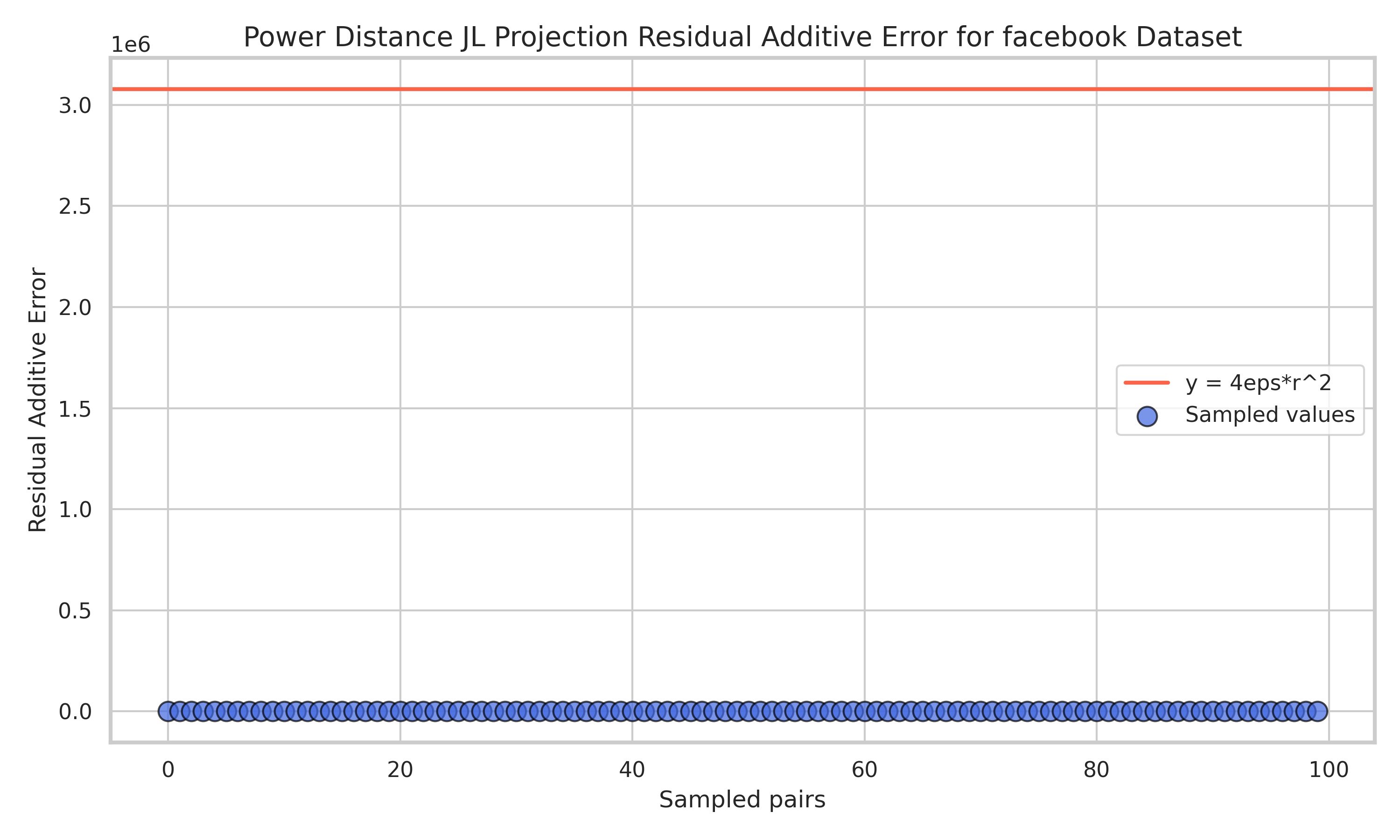}
        \caption{Residual error on facebook}
        \label{fig:sub4}
    \end{subfigure}
    
    \caption{Illustrations of Power Distance JL Transform Residual Error}
    \label{fig:four_figures}
\end{figure}

\begin{figure}[h]
    \centering
    \begin{subfigure}{0.48\linewidth}
        \centering
        \includegraphics[width=\linewidth]{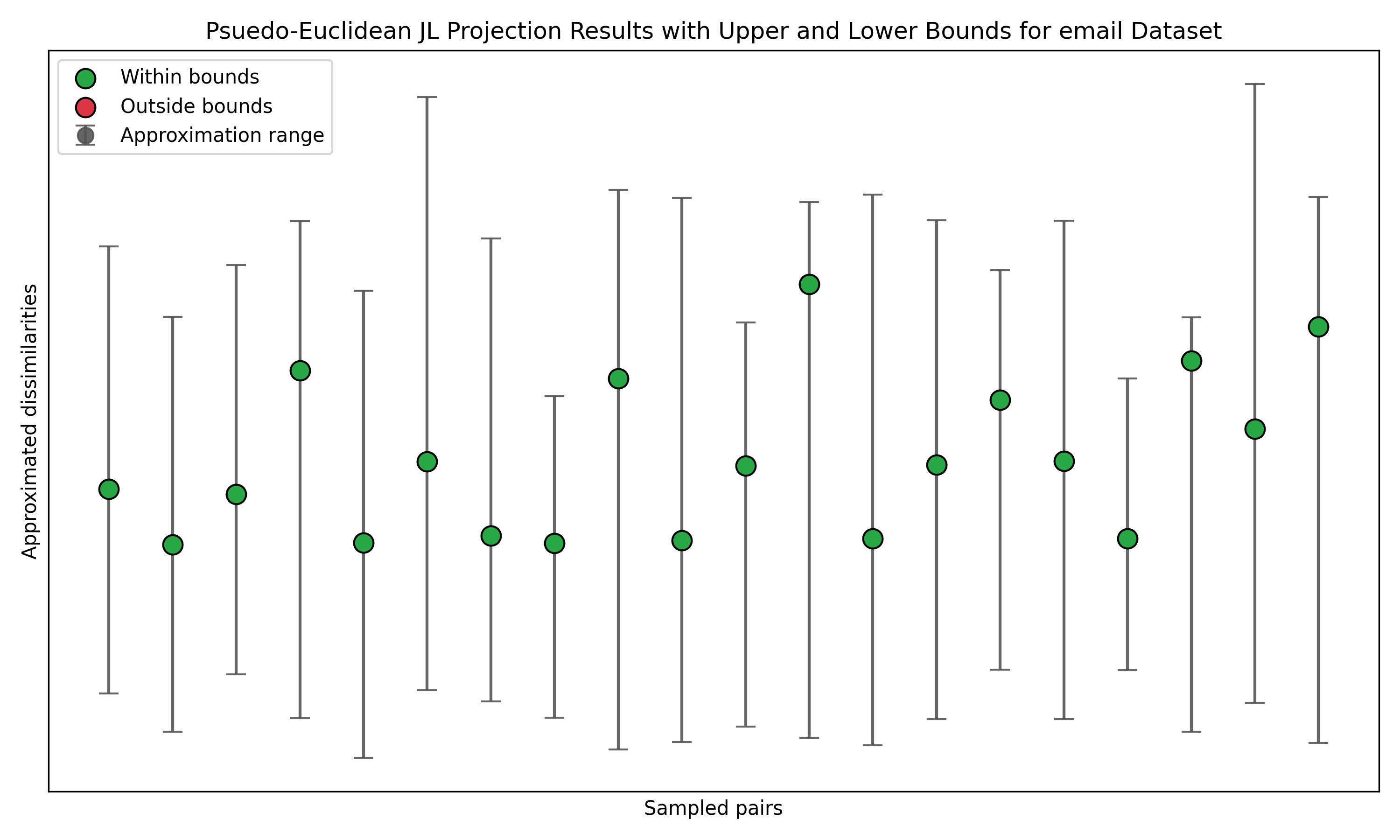}
        \caption{Approximation factor for email}
        \label{fig:sub1}
    \end{subfigure}
    \hfill
    \begin{subfigure}{0.48\linewidth}
        \centering
        \includegraphics[width=\linewidth]{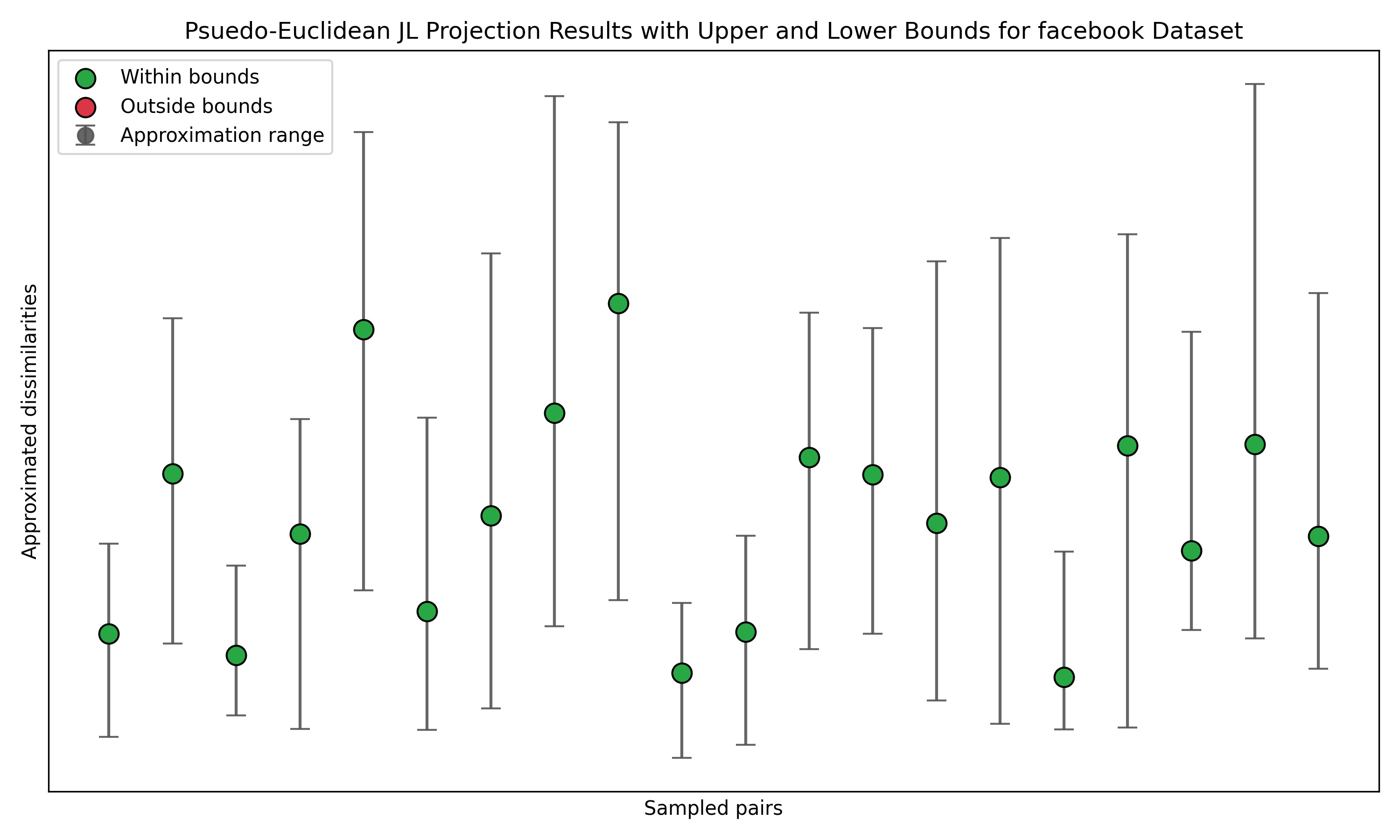}
        \caption{Approximation factor for facebook}
        \label{fig:sub2}
    \end{subfigure}
    
    \vspace{0.5em}
    
    \begin{subfigure}{0.48\linewidth}
        \centering
        \includegraphics[width=\linewidth]{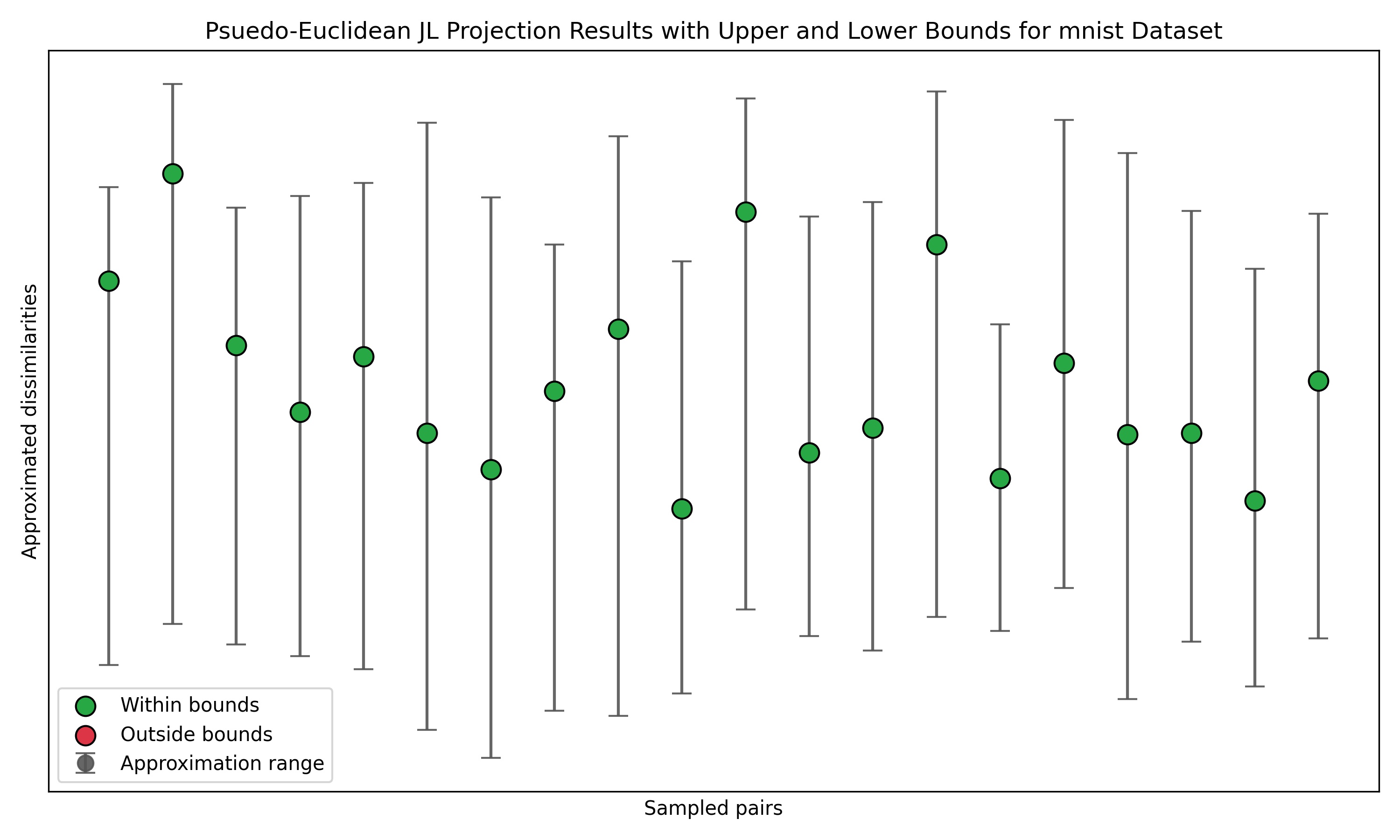}
        \caption{Approximation factor for MNIST}
        \label{fig:sub3}
    \end{subfigure}
    \hfill
    \begin{subfigure}{0.48\linewidth}
        \centering
        \includegraphics[width=\linewidth]{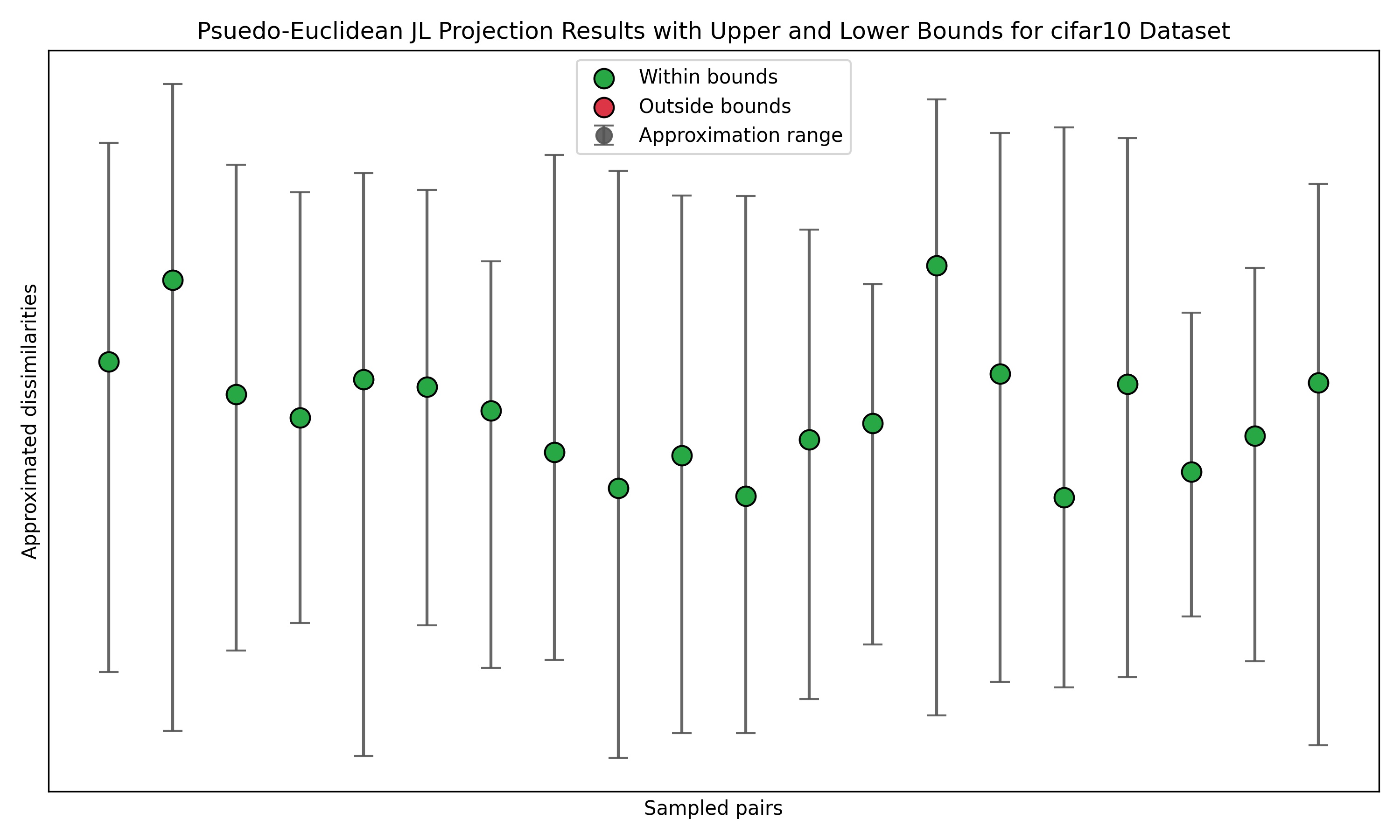}
        \caption{Approximation factor for CIFAR-10}
        \label{fig:sub4}
    \end{subfigure}
    
    \caption{Illustrations of Pseudo Euclidean JL Transform Multiplicative Error}
    \label{fig:four_figures}
\end{figure}

\end{document}